\documentclass[12pt,english]{article}
\usepackage[T1]{fontenc}
\usepackage[latin9]{inputenc}
\usepackage{geometry}
\usepackage{amstext}
\usepackage{amssymb}
\usepackage{babel}
\usepackage{color}
\usepackage{amsmath}
\usepackage{graphicx}
\usepackage{epstopdf,epsfig}
\usepackage{float}
\usepackage{color}
\definecolor{newredcomments}{RGB}{170,23,23}

\definecolor{newred}{RGB}{144,23,23}

\usepackage{tgpagella} % delete this to change back to the default font
\usepackage[toc]{appendix}
\usepackage[numbers]{natbib}
\setcitestyle{authoryear,open={(},close={)}}
\usepackage{threeparttable}
\usepackage{rotating}

\usepackage{booktabs,caption}

\providecommand{\keywords}[1]
{
	\small	
	\textbf{\textit{Keywords---}} #1
}

\providecommand{\JELcodes}[1]
{
	\small	
	\textbf{\textit{JEL:}} #1
}

\usepackage[unicode=true,
bookmarks=true,bookmarksnumbered=false,bookmarksopen=false,
breaklinks=true,pdfborder={0 0 0},backref=false,colorlinks=true]
{hyperref}
\hypersetup{pdftitle={Microcredit},
pdfauthor={Celik, Khazanov, Moav, Neeman, Zoabi},
linkcolor=red, citecolor=newred, urlcolor=blue, filecolor=blue}

\begin{document}

\title{Why Not Borrow, Invest, and Escape Poverty?  
 \footnote{We are grateful to Gianni Marciante and Ilia Sorvachev for very helpful research assistance. We benefited from comments from: Arun Advani, Susanto Basu, Pablo Beker, Valery Charnavoki, James Choy, Diana Egerton-Warburton, Moshe Hazan, Dilip Mookherjee, Kirill Pogorelskiy, Debraj Ray,  Rajiv Sethi, Ofer Setty, Selma Walther, David Weiss and participants in the CAGE-AMES seminar at the University of Warwick.  Celik: City, University of London, Northampton Square, London EC1V 0HB, United Kingdom, email:  dagmara.celik@city.ac.uk. Khazanov: Department of Economics, The Hebrew University of Jerusalem, Jerusalem, Israel, email: alexey.khazanov@mail.huji.ac.il. Moav: Department of Economics,  University of Warwick, Coventry CV4 7AL, United Kingdom, email: o.moav@warwick.ac.uk. Neeman: Berglas School of Economics, Tel Aviv University, Tel Aviv, Israel, email: zvika@post.tau.ac.il. Zoabi: The New Economic School, 100 Novaya Street, Skolkovo Village, The Urals Business Center, Moscow, Russian Federation, email: hosny.zoabi@gmail.com. The data collection was funded by the City, University of London (Pump Priming grant no. 445258). AEARCT identification number is AEARCTR-0007461. }}
\author{
Dagmara Celik Katreniak\\
  \begin{minipage}[t]{3.3cm} \centering  \footnotesize City, University of London \end{minipage} \and
Alexey Khazanov\\
  \begin{minipage}[t]{3.3cm} \centering   \footnotesize The Hebrew University of Jerusalem \end{minipage} \and
Omer Moav\\
  \begin{minipage}[t]{3.3cm} \centering   \footnotesize University of Warwick and Reichman University\end{minipage} \and
Zvika Neeman\\
  \begin{minipage}[t]{3.3cm} \centering   \footnotesize Tel Aviv University \end{minipage} \and
Hosny Zoabi\\
  \begin{minipage}[t]{3.3cm} \centering   \footnotesize The New Economic School\end{minipage}
}

\date{\today}

\maketitle

\bigskip

\begin{abstract} 
Take up of microcredit by the poor for investment in businesses or human capital turned out to be very low. We show that this could be explained by risk aversion, without relying on fixed costs or other forms of non-convexity in the technology, if the investment is aimed at increasing the probability of success. Under this framework, rational risk-averse agents choose corner solutions, unlike in the case of a risky investment with an exogenous probability of success. Our online experiment confirms our theoretical predictions about how agents' choices differ when facing the two types of investments.

\end{abstract}

\keywords{ Risk aversion, Microfinance, Investment, Poverty, Development, Inequality}

\JELcodes{O12 , I32}

\thispagestyle{empty}

\newpage

\section{Introduction}
The UN declared 2005 the ``Year of Microcredit,'' and the 2006 Nobel Peace Prize was awarded to Muhammad Yunus and the Grameen Bank for their contribution to reducing world poverty by providing affordable credit to the poor. There were high hopes, based on a large body of research and a widely held view, that alleviating credit constraints would allow the poor to make profitable investments in small businesses and escape poverty. Less than a decade later, research started to show that access to low-interest credit, provided by the thousands of microcredit NGOs that were established, did not deliver as anticipated. 

\cite{banerjee2015miracle} find, in a randomized evaluation in India, low take-up of microcredit; no increase in households that are business owners, and despite some increase in the size of existing businesses, the average business remained small and not very profitable. Low take up and limited impact on business creation is also found in other places.\footnote{See, for instance, \cite{Banerjee2013}, \cite{Crepon2015}, \cite{Angelucci2015}. \cite{Tarozzi2015} find higher take up in Ethiopia, but, similarly to other regions, no significant impact on business creation.} \cite{meager2019understanding}  jointly estimates the effect across seven studies and finds that the impact of microcredit on household business is unlikely to be transformative and may be negligible. In a subsequent meta-analysis, \cite{meager2022aggregating} finds a precise zero effect for poor households and uncertain yet large effects for wealthier households with business experience.\footnote{In addition, \cite{meager2022aggregating} discusses some harmful effects of microcredit and provides further references.}

What went wrong? What can explain the gap between the theory that raised such high hopes, and the disappointing realization? Perhaps, as argued by \cite{banerjee2015miracle}, there is simply less potential for high-return businesses for the poor than anticipated by microcredit enthusiasts, maybe because of a lack of complementary factors such as proper training. Evidence indicates, however, that the poor did succeed in business creation when they were given the productive assets or the cash to make the investment (e.g. \cite{doi:10.1093/qje/qjx003}, \cite{handa2018can}). So perhaps the poor can run a small business and will do so, if the risk associated with borrowing is removed.

Consistent with this, \cite{bd2011book} describe in much detail how reducing risk plays a crucial role in the lives of the poor, and \cite{Banarjee2000}, in anticipation of the disappointing outcomes of microcredit, proposes that the poor are too risk-averse to borrow and invest. In his model, an investment project is subject to a fixed cost, and it succeeds with an exogenous probability. The poor, despite high expected returns, do not invest because they are too vulnerable to absorb the risk of failure.
However, \cite{Kraay2014} survey the empirical literature and conclude that the evidence is inconsistent with technology-based (fixed costs or S-shape production function) poverty traps.\footnote{We discuss this literature in the next section. We note here, however, that following \cite{Kraay2014}, we do not claim that there are no businesses that require significant fixed costs. The claim is that the existence of business opportunities that do not require significant fixed costs is sufficient to conclude that the evidence is inconsistent with technology-based poverty traps. In other words, the puzzle is why don't the poor invest in these businesses. In addition, note that the existence of `s-shaped poverty dynamics' in the data (e.g., \cite{10.1093/qje/qjab045}) doesn't indicate non-convexities in production. This is illustrated by several models (see the discussion in the next section) and it is also an outcome of our model.}

To address this critique, \cite{bd2011book} propose that the poor do not borrow to invest, not even a small amount they could afford to risk, because they believe, in spite of the facts, that production functions are S-shaped. This proposition may be true, but an explanation that is not based on false beliefs might be preferred, as this would be consistent with the claim, also made by  \cite{bd2011book}, that the poor have a very good understanding of their economic environment. 

We propose an explanation that suffers from none of these limitations. We show that risk aversion can explain why the poor do not borrow, invest, and escape poverty when they have access to microfinance. Our explanation does not rely on an S-shape production function, fixed costs, or any other non-convexity in the technology, or any limited rationality such as false beliefs.  

We argue that the probability of success of a business isn't exogenous: investment is directed not just to increase the profit of the business if successful, but to a large extent also to increase the probability that it is successful. For instance, a farmer could invest in a larger plot of land, which would increase revenue if rain is sufficient and the harvest is successful, or invest in an irrigation system that would deliver a more reliable supply of water and increase the probability of success. Similarly, a farmer could invest in more livestock, to increase profit in a good year with no extreme weather shocks or disease, or invest in shelter or vaccination to protect the livestock and increase the probability of success. An owner of a restaurant or coffee shop could invest to increase space for clients to increase profits if the business is successful, or invest in advertising and service quality to increase the probability of success. Any small business selling services could invest in skills and equipment to increase expected revenue and the probability that the business succeeds. It is, in fact, hard to think of any business in which investment cannot be directed to increasing the probability of success, even if it also increases the return if the business is successful.\footnote{In our model, individuals borrow the funds to invest in their business. We do not formally model investment in time and effort, which to a significant extent could also increase the probability of success. However, since the opportunity cost of investing time and effort is forgone income, this is equivalent to investing capital. Note that for the sake of simplicity, it is often assumed in the existing literature that the probability of success is exogenous, but one shouldn't conclude that this is a claim about the real world. We are not the first, of course, to assume in a model that the probability of success is a function of investment, but rather the first to show that this could lead to corner solutions.}  

We show that when investment increases the probability that the business is successful, corner solutions could emerge even in the absence of S-shape production functions (or other non-convexities in the technology). In particular, we consider investment projects with a binary outcome of success or failure, where investment increases the probability of success at a constant (or diminishing) rate. We show that the expected utility of a risk-averse agent as a function of the size of investment is typically U-shaped in these projects, and as a result, it could lead the risk-averse to avoid any investment, despite the high expected return at low levels of investment. In addition, a decline in risk aversion leads to an increase in expected utility at high levels of investment. Thus, as consistent with the evidence, the poor, who are highly risk-averse, choose zero investment, whereas the wealthier, who are less risk-averse, choose to borrow and invest.  

To understand the logic of the U-shaped expected utility, consider a lottery with zero expected return: with probability $p$ the outcome of the lottery is a prize of one dollar, and with the complementary probability $1-p$, it is zero. The cost of generating a probability of success $p$ is equal to $p$ dollars. That is, with no investment the probability of winning the prize is zero; with investment of one dollar the probability is one; if the investment is half a dollar $p$ is one half, and so on. In this case, any individual is clearly indifferent between no investment and investing the maximum ($p=1$), where with probability one the agent simply receives the dollar invested back. The outcome is certain and identical in both cases. For any investment strictly between the two corners, the expected return is the same as for the two corners, but the realization is risky. Therefore, by definition of risk aversion, any risk-averse individual would strictly prefer the corners over any other investment strictly between zero and one.\footnote{Unfortunately, we cannot provide a complete answer to the question of how general our U-shape result is. We prove it holds for CARA utility functions. The intuition for our U-shaped expected utility doesn't depend on the specific form of the agent's preferences. In Appendix \ref{CRRA} we provide a numerical example that shows that the result extends also to the case of a CRRA function, albeit with an ``approximate'' rather than an exact U-shape. This suggests that our qualitative results hold more generally, albeit in a weaker form. In any case, the result that agents' investment is discontinuous in risk aversion still holds in the example, and may well hold more generally.} 

To understand why the poor might avoid a high expected return investment that the wealthy would invest in, consider two changes to the lottery. First, the reward in case of a successful outcome is greater than one, so that the expected return of the lottery is positive. Second, the investment is limited to be strictly below one, so that success cannot be guaranteed by high investment. Risk averse agents, who would typically choose between one of the two corners, now face a tradeoff between avoiding risk (by not investing) and enjoying an expected positive return (by investing the maximum possible). If the reward is not too high and risk aversion declines with wealth, there would be a wealth threshold above which individuals invest in the project and below which they don't.\footnote{In an extended model presented in Appendix \ref{App_Resource_Constrain}, we show that our main result doesn't depend on any constraints on borrowing. Moreover, even if agents could declare bankruptcy (in the case that they borrowed, were unsuccessful in their investment, and could not repay the entire debt), and liability is limited, the results still hold as long as: (1) The wealth of individuals after bankruptcy is correlated with their wealth before bankruptcy (e.g., they can hide some wealth), and (2) bankruptcy following an unsuccessful investment leaves individuals worse off, in comparison to the option of not investing.}

We believe that our simple result -- that risk aversion can lead to corner solutions, despite the absence of fixed costs or any other non-convexity -- may have been overlooked in the existing literature because of the conventional modeling of risky investment, where the probability of success is exogenous. If the agent decides how much to invest in an asset, but the investment has no effect on the probability of success, only on the reward if the investment is successful, the expected utility of a risk averse agent is a concave function of investment or an inverse U-shape, leading to an interior solution. The optimal investment increases with wealth (if risk aversion is declining with wealth), but the change is continuous.\footnote{In such a framework, the S-shape of wealth dynamics, required for a poverty trap, is achieved by assuming a fixed cost  (see, for instance, \cite{Acemoglu1997}, in addition to \cite{Banarjee2000} mentioned above). \cite{aghionbolton1997} model the probability of success as an endogenous outcome of investment, but they do not show our main result.}

Thus, one might wonder why the poor don't invest in businesses in which the probability of success is given, and the investment increases the size of the reward if the investment results in success. We propose that businesses that are available to the poor include both margins of investments. Poor individuals can invest to increase the reward in case of success, and they also have the option to invest in order to increase the probability of success -- and without any investment in that margin, the probability of success is rather low. We show in section \ref{Sec_TwoMargins} that, for the same expected return, a risk-averse agent will always prefer to invest in increasing the probability rather than in increasing the reward: investing in the probability second-order stochastically dominates investing in the reward. Thus, as long as a significant part of the return to investment is in the form of a higher probability of success, our results hold.

Investment in our model cannot be diversified among independent projects to eliminate the risk. This is, of course, standard in theories that are based on risk aversion. We propose that investment augments the expected productivity of an existing indivisible asset -- the individual's labor, which is, for poor people, their main productive asset. The investment could be in human capital that increases the probability of finding a well-paying job or investing in a small business in which the owner's labor is the main factor of production. In addition to the examples above (illustrating the two margins of investment), consider an individual whose business is to provide services such as plumbing or electricity, or even simple manual tasks. She could invest in augmenting relevant skills, health, and physical abilities; invest in marketing the services; purchase useful complementary tools, or spend money on anything else that increases the probability that the business is successful and profits are higher if successful. Other small businesses are not that different. The owner has her own time and could augment the expected income of the business with investment in complementary factors or intermediate goods, such as a larger stock and more space for storage in a shop, or more fertilizer, tools, and irrigation equipment in farming.\footnote{We are aware that the poor do diversify to reduce risk, but, of course, diversification is limited, and cannot remove risk altogether.}

We designed an online experiment to test the model's main predictions regarding the investment choices of risk-averse individuals. In particular, we tested the prediction that investment in increasing the probability of success leads to corner solutions, compared to investment in the reward for success when the probability of success is exogenous. We also tested the prediction that for the same expected return, risk-averse individuals would prefer to invest in increasing the probability of success rather than in the return if successful. Participants were recruited at random from a representative sample in the Czech Republic.

One might raise the concern that representative households from the Czech Republic are substantially better off than a typical microfinance borrower. However, the experiment was designed to test the prediction of the model regarding human behavior -- investment choices under risk -- regardless of its application to the question of the failure of microfinance. Thus, there wasn't a good reason to conduct the experiment with a sample of people from a developing country, where the ability to conduct online experiments is limited.\footnote{Due to the Covid-19 pandemic, we had to execute the experiment online.} Our experimental evidence supports our model's predictions at all income levels. Poor, risk-averse households in our sample choose to avoid high-return investment, unlike the wealthier households that do invest. Thus, if poor households in our sample, which are wealthier than the typical poor household in a poor country, do not invest, then poorer households - such as those targeted by microfinance - would also avoid such an investment. In reaching this conclusion, we follow the conventional claim that the poor in poor countries are not fundamentally different than other people, they simply have less income.

Participants in the experiment decided how much to invest in three different games, with an endowment of 150 CZK ($\approx$ 5.75 Euro at the time of the experiment) in each of the games. In the ``probability game'' the reward for winning was 270 CZK. The probability of winning the prize was a function of the participant's investment. The higher the investment, the higher the probability of winning the reward. Participants lost the entire sum invested with the complementary probability. The probability as a function of investment was set such that the expected return of the probability game was constant at 50\%. In the ``reward game'', the prize was determined by the decision of the participant: it was equal to three times the investment. The probability of winning the prize was set to 50\%, as was the probability of losing the sum invested. Note that the expected return of the reward game was constant at 50\%, as it was for the probability game.\footnote{The stakes might seem too low for risk aversion to be at play. But for the poor households in our sample, this isn't the case. Poor households, on average, have to work nearly two and a half hours to earn 270 CZK, and it is about 75 minutes for households with average income.}

Consistent with the model's predictions, when we limit the analysis to the risk-averse participants who understood the game, the results of the two games show that participants tend to choose the corners in the probability game significantly more than in the reward game. In particular, in the reward game, the distribution of investment was unimodal  (with a mean of 81 CZK and a median of 90 CZK). In the probability game, the distribution of the investment (by the same participants) was bimodal with significantly more corner (or near corner) investment decisions (0 or 30 CZK and 120 or 150 CZK). 

The third game was a ``step-by-step game.'' 30 CZK of each participant's endowment, of the 150 CZK, were invested by the experimenter on behalf of the participant, in a lottery in which they can win 270 CZK with a probability of 1/6 and zero otherwise. The participants were then asked to make four subsequent decisions about how to invest each additional 30 CZK. In each step, they had the option to direct investment towards increasing the probability of success, or to increase the reward they receive if successful. Rewards and probabilities were set such that the expected return was constant at 50\% in each of the stages for both investment options.

Consistent with the outcome that investment in increasing the probability second-order stochastically dominates investment in increasing the reward, when the risk-averse participants could choose between the two margins of investments they opted to increase the probability of their success, rather than increasing the reward, in the majority of junctions. Specifically, slightly more than 70\% of the risk-averse participants invested in increasing the probability in at least three of the four steps, whereas slightly less than 10\% invested in increasing the reward in at least three steps. The remaining risk-averse participants (slightly less 20\%), split their investment equally between the two margins.

Our main contribution is to propose a theory that could explain why the poor tend to turn down the opportunity to borrow in order to invest in high-return projects. The theory is based on risk aversion, which is central in the life of the poor, without relying on fixed costs or any other non-convexity in the production technology. We discuss the related literature, including the evidence on the importance of risk aversion in the lives of the poor, and the evidence suggesting that fixed costs are negligible in many investment opportunities facing the poor, in the next section. We postpone the discussion of policy implications to the concluding section, after presenting the model and the experiment.

\section{Related Literature}

A key assumption of our theory is that the poor are risk-averse and that reducing exposure to risk is central in their lives. The facts seem consistent. \cite{bd2011book} argue that "[r]isk is a central fact of life for the poor, who often run small businesses or farms ... with no assurance of regular employment. In such lives a bad break can have disastrous consequences" (p. 133). They further argue that the poor are constantly worrying about the future, particularly about imminent disasters, and take a variety of ingenious and costly precautionary measures to limit the risks they are exposed to, such as managing their businesses conservatively and diversifying their portfolio of activities, including by marriage and temporarily migration (pp. 141 - 143). Consistent with this, \cite{Morduch1990} presents evidence suggesting that the poor avoid profitable but risky technologies. \cite{Bryan2014} show that the poor avoid low-cost seasonal migration that is highly rewarding because of the fear of failure, and \cite{Karlan2014} and \cite{Cole2017} show that insurance against adverse weather shocks can induce farmers to invest more in high-return risky production options.\footnote{Our model assumes that full insurance against risk (at an affordable cost) isn't available. We believe that this is a reasonable assumption justified by moral hazard and adverse selection problems. \cite{Udry1990}, \cite{Townsend1995}, and \cite{Morduch1995}  provide evidence that the poor are often insured against the risks they take, but as suggested by \cite{Banarjee2000}, these studies only observe the risky activities people have chosen to take. The poor may have foregone other investment opportunities to limit the risk they bear. Moreover, \cite{Townsend1995} shows that full insurance is limited to some risks. In addition, the \cite{wbreport2001} reports that ``poor people, even though they need insurance most, are more likely to drop out of informal [insurance] arrangements.'' (p. 144). Finally, \cite{bd2011book} show that the poor avoid insurance, in particular health insurance, because ``[c]redibility is always a problem with insurance products.'' (p.153).}

Risk aversion declines with income and is particularly significant among those who live in extreme poverty -- the same people who are a major target of poverty-reduction policies, such as microfinance (\cite{andrisani1978}, \cite{hill1985}, \cite{Cicchetti1994}, and \cite{Shaw1996}.) The \cite{wbreport2001} report shows that "[the] poor are highly risk averse and reluctant to engage in the high-risk, high-return activities that could lift them out of poverty" (p. 138), and "[a]s households move closer to extreme poverty and destitution, they become very risk averse" (p. 145). Finally, \cite{bd2011book} argue, as consistent with the structure of our model, that investment is often equivalent to buying a lottery ticket (p. 87). For instance, the outcome of schooling is employment by the government or a large firm, if successful, or subsistence self-employment, if not. 

Much of the earlier literature that offers explanations for the persistence of poverty abstracts from risk aversion. This literature proposes theories based on credit constraints combined with increasing marginal returns to investment, such as fixed costs or S-shape production functions.\footnote{E.g., \cite{Dasgupta1987}, \cite{Galor1993}, \cite{Banerjee1993}, \cite{benabou1996equity}, \cite{durlauf1996theory}, \cite{ghatak2002simple}, \cite{mookherjee2003persistent}, and \cite{mookherjee2010aspirations}, among many others.} In these models, the poor cannot benefit from high returns on the modest investments they can make, and as a result they cannot gradually escape poverty. One significant limitation of these papers, which our model resolves, is that they do not explain why the poor leave high return investment opportunities unexploited when they do have access to credit, which is what we observed when microcredit became available. 

One can solve this puzzle by introducing risk to these models that are based on fixed costs or other non-convexities (as in Banerjee (2000) discussed in the introduction). The risk-averse poor might choose the safe alternative and avoid the investment, even if credit is available. The downside outcome -- if the investment fails to yield the expected high return -- could be too painful, leaving the poor with a low income, fewer assets, and debt. In this theory, the fixed cost plays a crucial role, as it prevents the poor from investing a modest amount that exposes them to a level of risk they are willing to accept.\footnote{The argument that risk aversion leads to under-investment isn't new, of course. It is proposed by \cite{Stiglitz1969} and is further developed, with an emphasis on the poor, by many others (See the literature review in \cite{Banarjee2000}).}

However, \cite{bd2011book} claim that many investment opportunities, such as in education or health, offer a high expected return with no significant fixed cost. They argue that the marginal return to investment in education is high at low levels of investment: "... every little bit of education helps ... People who go on to secondary education are more likely to get a formal-sector job, but even those who don't are able to run their businesses better" (p. 82). \cite{Kraay2014} survey the empirical literature and, as mentioned in the introduction, conclude that the evidence is inconsistent with technology-based poverty traps.\footnote{Not much capital is needed to start a business in a developing country and returns to investment are very high:  5\% to 20\% per month, at investment levels as low as 100 dollars (\cite{mckenzie2006entry}, \cite{Mel2008, mel2009, Mel2012}, \cite{Fafchamps2014}). Similarly, despite high returns, many farmers decline to invest in fertilizer that is available in small quantities \cite{Duflo2011}), and many shopkeepers fail to make small inventory investments (\cite{Kremer2011}).} 

Our model succeeds in explaining the disappointing outcomes of microcredit without relying on fixed costs (or other non-convexities in the technology).\footnote{\cite{piketty1997dynamics} advocates the importance of removing fixed costs from models of persistence of poverty. In his introduction he writes: ``[the existing model's] mechanism is different from ours in that it relies entirely on a non-convex technology.'' However, he assumes a non-convex technology in effort rather than in capital.} We are not the first to remove increasing marginal productivity from a poverty trap model. The existing literature, however, focuses on various explanations for low investment by the poor in the absence of available credit.\footnote{This literature includes, among others, \cite{Moav2002} who shows that if the marginal propensity to save increases with income a poverty trap could emerge; \cite{Chakraborty2005} and \cite{Moav2005} obtain similar results based on the interaction between health and human capital in the former and the trade-off between fertility and education in the latter; \cite{bm2010workingpaper} and \cite{Bernheim2015} focus on "temptation goods" and self-control problems, and \citeauthor{Moav2010} (\citeyear{Moav2010,Moav2012}) focus on conspicuous consumption as reasons for low saving by the poor.} We offer an explanation for why the poor do not invest when they do have access to credit.

There are only a few other explanations for why the poor avoid small affordable investments with high expected returns. \cite{bd2011book}, as mentioned in the introduction, suggest that the poor typically believe, in spite of the facts, that the production function has an "S-shape" -- in order to enjoy a high return, the investment should be large. The combination of false beliefs and the risk associated with the investment push the poor to avoid it all together. "In reality, there should not be an education-based poverty trap: Education is valuable at every level. But the fact that parents believe that education is S-shaped [...] create[s] one." (p. 89). This claim is supported by some evidence (e.g., \cite{Nguyen2008}) but other evidence presented by the authors led them to conclude that the poor have a good understanding of their economic environment: "the poor are no less rational than anyone else -- quite the contrary. Precisely because they have so little, we often find them putting much careful thought into their choices: They have to be sophisticated economists just to survive." (p. ix). Our model neither requires false beliefs about reality nor any other ``behavioural'' elements. 

\cite{Kremer2013} argue that poor households in developing countries reveal risk aversion in small-stakes gambles that cannot be explained by any reasonable degree of risk aversion within expected utility theory, and propose that loss aversion within prospect theory may play a role. Our simple calibration, based on parameter values estimated from \cite{AugsburgDeHaasHarmgartMeghir15}, demonstrates that within our model, reasonable values of the risk aversion coefficient are sufficient to prevent the poor from investing, without relying on loss aversion (Appendix \ref{Sec_SimpleCalibration}).

More recently, \cite{banerjee2015miracle}, also mentioned in the introduction, propose that the low take-up of loans for starting a business could be an outcome of the lack of complementary factors such as proper training or skills, and more generally, that there is less potential for high return businesses for the poor than anticipated by microcredit enthusiasts. However, \cite{doi:10.1093/qje/qjx003} show that when poor women receive a productive asset (a couple of cows) and some relevant training, they have the skills to run a simple yet successful business that alleviates poverty in the long run. These results seem consistent with our theory. Women who could borrow the funds for the required investment and training avoid it, despite the high return.  Similarly, \cite{handa2018can} find significant effects of cash transfer programs on productivity. The main difference between (a) borrowing for investment and (b) being given the asset -- or the cash to purchase a productive asset with no debt -- may be risk.

%{\color{red}
%Handa, S., Natali, L., Seidenfeld, D., Tembo, G., & Davis, B. (2018a). Can unconditional cash transfers raise long-term living standards? Evidence from Zambia. Journal of Development Economics, 133, 42–65.}

\section{Theory \label{sec:section2} }

In this section we present our theoretical model and derive the following three main results: 

\begin{enumerate}
    \item In the \emph{probability model} -- the model in which an agent's investment increases the probability of success -- the expected utility of an agent with a CARA utility function is U-shaped in the agent's investment. Consequently, investment is \emph{discontinuous} in the agent's level of risk-aversion: more risk averse agents invest nothing, while less risk averse agents invest the largest amount possible.
    \item In the \emph{reward model} -- the model in which an agent's investment increases the reward for success -- the expected utility of an agent with any monotone increasing  utility function is concave (inverse U-shaped) in the agent's investment. It follows that the agent's investment is \emph{continuous} in the degree of risk-aversion.
    \item If investment in the  probability of success and investment in the reward obtained upon success are both available and generate the same expected income, then a risk averse agent prefers to invest in increasing the probability.
\end{enumerate}
 
Our theoretical results have three main predictions regarding optimal investment choices of risk averse individuals. We examine the relevance of the theoretical results with an experiment.

\begin{enumerate}
    \item The investment choices of individuals in the \emph{probability game} -- the experimental exercise in which an agent's investment increases the probability of success -- give rise to a bimodal distribution.
    \item The investment choices of individuals in the \emph{reward game} -- the experimental exercise in which an agent's investment increases the reward for success -- give rise to a  unimodal distribution.
    \item In the \emph{step-by-step game} there are two binary lotteries with identical expected payments that yield identical payments upon failure. Individuals can choose between investing in the probability of success or investing in a higher reward upon success. In this step-by-step game, individuals invest in the probability of success.
    
\end{enumerate} 

\subsection{The Probability Model}

An expected-utility-maximizing agent may invest in a project with binary outcomes: investment will be successful and generate a high return $H$, or it will fail, and generate a low return, $L<H$.

In the probability model, the agent controls the project's probability of success, $p\in[0,\bar{p}]$, at a linear cost $c(p)=\alpha p$ for some $\alpha>0$. Importantly, we assume that $\bar{p}<1$, so that, no matter how much the agent chooses to invest, they cannot ensure that investment would be successful. We assume that investment generates a positive expected return for the agent. That is, $pH+(1-p)L - \alpha p > L $ for any $p>0$, or equivalently, $\alpha<H-L$.

\bigskip 
\noindent \textbf{Proposition 1.} \emph{The expected utility from investment in the probability model of an agent with a CARA utility function $u(x)=-e^{-\lambda x}$ is U-shaped in the agent's level of investment $p$. That is, there exists some threshold $\hat{p}\in[0,1]$ such that the agent's expected utility is decreasing in $p$ on the interval $[0,\hat{p}]$ and increasing in $p$ on the interval $[\hat{p},\bar{p}]$.}
\bigskip

% PROPOSITION FROM APPENDIX VERSION
%\noindent \textbf{Proposition 1.} \emph{The agent's expected utility function $U(p)$ is U-shaped in $p$. Namely, it is non-increasing on an interval $[0,\hat{p}]$ and increasing on the interval $[\hat{p},\overline{p}]$ for some $\hat{p}\in[0,\overline{p}]$. Notice that $U(p)$ may be either nonincreasing or nondecreasing on the entire range. }

\bigskip

The parameter $\lambda>0$ in a CARA utility function describes the agent's Arrow-Pratt coefficient of risk aversion. A higher value of $\lambda$ indicates a higher degree of risk aversion, and as $\lambda$ decreases to zero, preferences converge to risk neutrality. It therefore follows that:

\bigskip

\noindent \textbf{Proposition 2.} \emph{The optimal choice of the level of investment $p$ of an agent with a CARA utility function with parameter $\lambda$ is discontinuous in $\lambda$. There is a threshold level of risk aversion $\lambda^{o}>0$ such that more risk averse agents choose the minimum investment $p=0$, and less risk averse agents choose the maximum investment $p=\overline{p}$.}

\bigskip

\bigskip

Unfortunately, we do not have a full answer to the question of whether the results described in Propositions 1 and 2 above generalize to other utility functions. Notably, the intuition for our U-shape result that we described in the introduction doesn't depend on the specific form of the agent's preferences. In  Appendix \ref{CRRA} we provide a numerical example that shows that our results extend also to the case of a CRRA function, albeit with an ``approximate'' rather than an exact U-shape, as is the case for CARA functions. This suggests that our qualitative results hold more generally, albeit in a weaker form. In any case, the result that the agent's investment is discontinuous in their level of risk aversion holds in the example, and may hold more generally.

\subsection{The Reward Model}

As in the probability model, an expected-utility-maximizing agent may invest in a project with binary outcomes: investment may succeed and generate a high return, or fail, and generate a low return.

In the reward model, the agent controls the project's return upon success rather than the probability of success, which we fix at $p$. As in the probability model, a failed project yields a payoff of $L$. A successful project yields a payoff of $H(c)\geq L$, where $H(c)$ is assumed to be increasing and weakly concave in the agent's cost of investment $c\geq0$. In this version of the model, the objective of the agent is to choose the cost of investment $c$ to maximize expected utility, which is not necessarily CARA.

\bigskip

\noindent \textbf{Proposition 3.} \emph{The agent's expected utility is concave in the cost of investment $c$. It follows that it is inverse $U$-shaped in $c$.}

\bigskip

As in the probability model, the agent's choice of level of investment is still decreasing in its level of risk aversion, but unlike in the probability model it is \emph{continuous} in the level of risk aversion.\footnote{Specifically, if $\{u_n\}$ is a sequence of utility functions that converges to a utility function $u$ and $\{c_n\}$ and $c$ are the associated costs of investment, then if $u_n$ exhibits more/less risk aversion than $u_{n+1}$ then $c_n$ is smaller/larger than $c_{n+1}$ and the sequence $\{c_n\}$ converges to $c$.}

\subsection{Probability vs. Reward} \label{Sec_TwoMargins}

In many situations, an agent faces a binary investment prospect in which they could invest either to increase the probability of success of a project, or invest to increase the reward upon success - or both. Which would the agent prefer? 

We consider a hybrid of our probability and reward models in which the agent's expected return from investment is held fixed, regardless of whether the agent invests in probability or in reward.

Our agent faces a binary project. The project succeeds with probability $p\geq p_0>0$ and yields $H(p)$, or it fails and yields $L$, where $H(p)>L\geq0$ for all $p\geq p_0$.

Suppose that the expected return of the project is constant so that $pH(p) + (1-p)L = C > L $ for every choice of $p\in [p_0,\bar{p}]$. The agent chooses whether to invest in the probability $p$ with the expected reward upon success of $H(p)=\frac{C-(1-p)L}{p}$, which is decreasing in $p$, or invest in the reward $H(p)$ with a declining probability of success, $p$.

We compare the $p$-lottery in which the agent receives $H(p)$ with probability $p$ and $L$ with probability $1-p$ with the $p'$-lottery, where $p > p'\geq p_0$, in which the agent receives $H(p') > H(p)$ with probability $p'$ and $L$ with probability $1-p'$. Our assumption that $pH(p) + (1-p)L \equiv C > L $ is fixed implies that these two lotteries return the same expected income to the agent.

\bigskip

\noindent \textbf{Proposition 4.} \emph{Suppose that $p > p'\geq p_0$. A $p$-lottery that pays $H(p)$ and $L$ with probabilities $p$ and $1-p$, respectively, generates a higher expected utility for a risk-averse individual than a $p'$-lottery that pays $H(p') > H(p)$ and $L$ with probabilities $p'$ and $1-p'$, respectively.}

\bigskip

\bigskip

It follows that under the circumstances described in this section, a risk-averse individual would prefer to invest in the probability of success than invest in the reward upon success.\\

\section{Empirical examination}

We tested our theoretical predictions using an online field experiment, drawing our participants from the Czech Republic. Participants answered a set of questions and made four incentivized decisions. The first decision elicited risk aversion. The next three decisions were investment decisions corresponding to the three main predictions of our theory: investment in the probability of success, to test the prediction of corner decisions; investment in a given probability of success, to test the prediction of interior decisions, and a choice between investing in increasing the probability of success or investing in increasing the reward if successful, to test the prediction that risk-averse individuals prefer to invest in the former if the expected return is equal in both options. The outcome of one of the decisions was randomly selected for payment.

\subsection{Experimental Design }

In the first incentivized decision (Decision 1), the participants were presented with a portfolio of 11 choices between a lottery and a safe option (inspired by \cite{Dohmen2010}). In the lottery, they could either win 1,300 CZK ($\approx$ 49.8 Euro) with 50\% chance or 0 CZK otherwise. The amount in the safe option increased in each row by increments of 100 CZK from 0 CZK up to 1,000 CZK (see Appendix Table \ref{AppTab1:RiskElicitation}). The participants were asked to select the first row in which they preferred the safe option over the lottery. The switching point is assumed to capture individuals' preferences toward risk. 

\begin{figure}[H] 
\begin{center}
\includegraphics[scale=0.35]{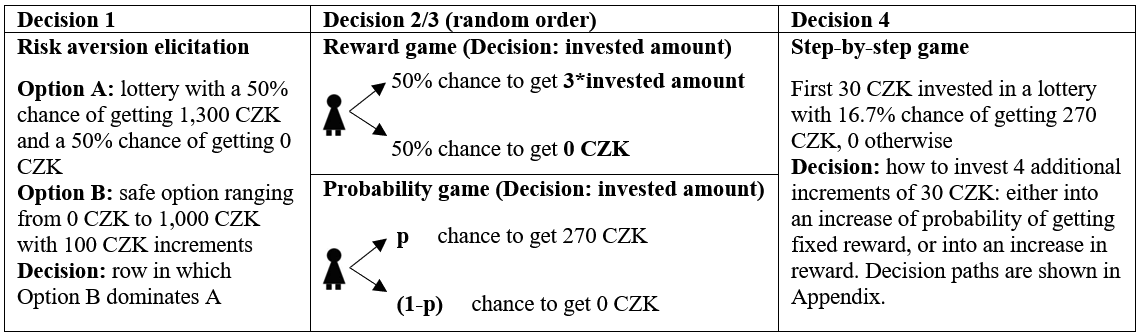}
\end{center}
\caption{\label{Fig:fig1}Experimental decisions}
\end{figure}

The remaining three incentivized decisions correspond to the investment decisions in three investment games: (i) {the ``reward game''; (ii) the ``the probability game''; and (iii) ``the step-by-step investment'' (for a visual summary of the games see Figure \ref{Fig:fig1}}). The games were designed to have the same expected returns. For each game, the participants were endowed with 150 CZK ($\approx$ 5.75 Euro) and were given detailed instructions, followed by a set of control questions to check the level of their understanding. For every correct answer, the participants received 5 CZK. If a participant answered any of the control questions incorrectly, we repeated the instructions and asked a different set of control questions (in which case the first set of questions would not be payment relevant anymore). Summary tables with all of the decision options were always displayed on participants' screens (see Appendix Figures \ref{AppFig1:ProbGameTable} and \ref{AppFig2:RewGameTable}). 

In the ``reward game'' the probability of success of an investment project is given. Participants invested part or all of their endowment. The success of their investment was determined by a roll of a fair die. Instead of dots, our die had a star on three sides and three blank sides. Thus, the participants had a 50\% chance of rolling a star, in which case their investment tripled. With the complementary probability of 50\% they received 0. The expected rate of return is therefore 50\%. 

In the "probability game", there was a fixed reward of 270 CZK ($\approx$ 10.3 Euro) upon winning, and the participants could determine their probability of winning.{\footnote{270 CZK is approximately equal to 2.4 hours of work in the two poorest income deciles and 1.27 hours in an average income household.} Every 30 CZK invested added a star on their die and increased their chance of getting the reward by 16.7\%.  In particular, there were five dice available with 1, 2, 3, 4, or 5 stars representing a 16.7\%, 33.3\%, 50\%, 66.7\% or 83.3\% chance of winning. Using a (non-standard) die ensured that participants understood which die had a higher chance of winning, and had a sense of what the probabilities are, even if they did not (or could not) calculate the exact probability of winning. The expected rate of return is 50\%, as in the ``reward game.'' The participants kept the non-invested money with certainty. The order of the reward and the probability games was randomized. 

Finally, in the ``step-by-step game'' the participants were informed that the first fraction (30 CZK) of their endowment was invested for them such that they had a 16.7\% chance of getting 270 CZK and an 83.3\% chance of getting 0. The participants were asked to make four sequential investment decisions. In each step, they were asked to invest an additional 30 CZK either into increasing the reward (keeping the probability of winning the reward fixed) or in increasing the probability of winning by 16.7\% (while keeping the reward fixed). All pathways are displayed in Appendix Table \ref{AppTab2:StepByStep}. The probabilities and rewards were set to ensure that at each step expected returns were equal. After the participants revealed their decision paths, they made their last payoff-relevant investment decision and decided how much of 150 CZK they would like to invest, where the probabilities and rewards were taken from their decision path. 

After the final decision and the last set of questions, we let the participants see the outcomes of all four incentivized decisions before one of them was selected for payment. First, the participants let the computer select one of 11 rows that would determine the outcome of their Decision 1 in the risk elicitation task. If the participants chose a lottery in the selected row, they would play the lottery: win 1,300 CZK with 50\% chance and 0 otherwise. If the participant chose a safe option instead, they would be reminded of the amount they would definitely receive if this decision was selected for payment. Then the participants were reminded of their investment decisions in the other three games (following the order they played the games). The computer displayed the invested amount, the probability of rolling a star, the expected reward, and the non-invested amount they were definitely receiving. The participants then rolled the digital dice and were presented with the outcomes of their decisions in the three investment games (decisions 2, 3, and 4). In the final step, the participants prompted the computer to determine which of the decisions would be selected for payment. The first decision was selected with 10\% probability, and the investment decisions 2, 3, and 4 were selected with 30\% probability each.\footnote{For a small sample of 102 participants Decision 1 was not incentivized. These participants are not significantly different from the other participants in terms of the observable characteristics and the two distributions of the decisions are similar (p-value of Kolmogorov-Smirnoff test equals 0.423). Given that the measure of risk aversion is essential for presenting the results, we decided to incentivize the risk elicitation among the rest of the participants, and we assigned this decision only a 10\% chance of being selected for payment.} All the instructions were clearly communicated to the participants prior to the game and reiterated in the outcome summary. The experimental protocol is available at \href{https://drive.google.com/file/d/1IDe4OMgMR6u-2pwQnGzI3F3A9fHBXwmd/view}{Online Appendix}.

\subsection{Final sample and randomization balance} 

The experiment was conducted in cooperation with IPSOS, a global market research company. Our sample comprises 846 individuals randomly selected from a representative sample of more than 22,000 people living in the Czech Republic. It took the participants on average 27 minutes to fill out the survey. Around 47\% of our sample is female. The average respondent is 43.5 years old, earns on average 19,789.6 CZK, and lives in a household with 35,484.6 CZK net income per month. These statistics are similar to the average net household and personal income per month in 2020 \citep{theoffice2020}. About 9.1\% of the respondents indicated primary education as their highest achieved education level. For 34.1\%, secondary school with a school graduation examination was the highest education level, 38.6\% finished vocational school, and 18.2\% received university-level education. About half of the sample is married and 36.2\% have at least one child living in the household with them. Frequencies of the responses in the various categories (the region they live in, the number of inhabitants, marital status, employment type, etc.) are similar for those who played the return game first and those who played the probability game first. Sample descriptive statistics and randomization balance checks are presented in Table \ref{AppTab3:DescrStat} in the Appendix.

Understanding the games is crucial for the interpretation of our results. If the participants do not understand the instructions, we cannot be sure whether their decisions represent their preferences or were due to a lack of understanding. While 49.7\% of participants understood both games fully (they answered all of the control questions correctly), 68.0\% understood them well (they made at most one mistake in answering the questions). In our analysis, we focus on the within-subject analysis of people with good/full understanding. We can rule out order effects given that the order of the games was randomized, and the randomization balance confirms that the participants who played the probability game first are on average the same as the participants who played the reward game first.

A crucial variable for our analysis is the participant's risk preferences. We used two measures of risk aversion, following closely the rich literature on the elicitation of risk preferences (e.g., \cite{Dohmen2010}). First, we asked participants a general question to rate their willingness to take risks on a 10-point scale (a general question is often used to elicit risk preferences and has been found to predict risk-taking behavior well \citep{Dohmen2010}). In our sample, 59\% ranked themselves below 5 on the 10-point scale and they would be classified as risk-averse. The average and modal response (5.2 and 5 respectively) and the distribution of responses are very similar to the distribution among a representative sample from Germany analyzed by \cite{Dohmen2011}. Second, as mentioned above, we asked the participants to look at eleven options in a lottery, in which they had a 50\% chance of winning 1,300 CZK and a 50\% chance of winning 0 CZK and a safe option which ranged from 0 to 1,000 with increments of 100 CZK. A risk-neutral individual prefers a lottery in rows 1 to 7 and a safe option in rows 8 to 11. Individuals are considered risk-averse if they switch above row 7, and not risk-averse if they do not. According to this measure, 74.8\% of the participants are risk averse, 6.7\% are risk-neutral, and 18.5\% are risk-loving. Similar proportions (78\%, 13\%, and 9\%) were found among the representative sample of adults from Germany \citep{Dohmen2010}. In our analysis, we present the results using the first measure but we replicate and present all the results with respect to the second risk-aversion measure in  Appendix Figures \ref{AppFig7:FMM_RAdohmen}, \ref{AppFig8:FMM_RAdohmen_sw} and Appendix Table \ref{AppTab7:probit_diffRAmeasures}.

\subsection{Results} 

To test our first theoretical prediction, we look primarily at the differences in the investment choices in the probability versus reward game. In this analysis, we use a within-subject design, and hence for each individual, we observe their decisions in both the probability game and the reward game. Our theory predicts that we would see higher dispersion of investments in the probability game compared to the reward game for the risk-averse participants (prediction 1). This is because in the probability game, we expect risk-averse participants to invest either (close to) the minimum or maximum amount from their endowment, resulting in a bimodal distribution of the investment choices. In the reward game, we expect the risk-averse participants to choose to invest some interior positive amount, resulting in a unimodal distribution of the investment choices (prediction 2). 

\begin{figure}[H] 
\begin{center}
 \includegraphics[scale=0.35]{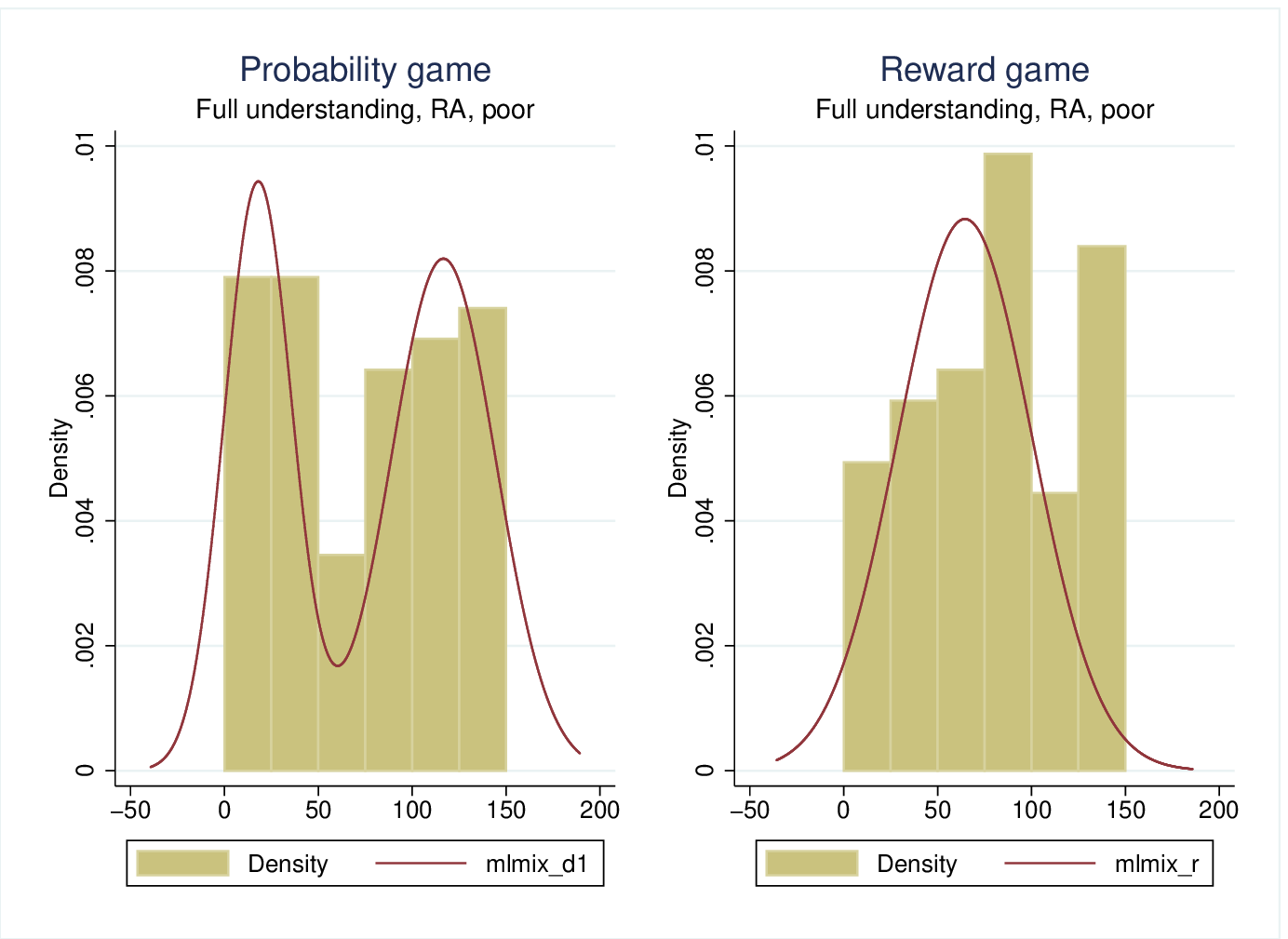}
\end{center}
      \begin{minipage}{\linewidth}\footnotesize 
      
      Note: Histogram of investment decisions in the probability and reward game for poor risk-averse participants. Overlaid is the distribution predicted by the Finite Mixture Models.
      \end{minipage}
\caption{\label{fig2} The comparison of investment decisions in the probability and return game }
\end{figure}

In Figure \ref{fig2} we visually compare the decisions of the participants in the probability game (left) and the reward game (right). On top of the histograms, we overlaid finite mixture models to model the distributions of choices in the two games. In line with our predictions, we find a bimodal distribution of the investment choices in the probability game, and a unimodal distribution in the reward game. The patterns are even more pronounced if we exclude people who invested the same amount across all three games (shown in Appendix Figures \ref{AppFig6:FMM_nonsw_full} and \ref{AppFig8:FMM_RAdohmen_sw}).\footnote{Some people have a tendency to stick to one number across multiple investment decisions, perhaps because they want to ensure consistency in their choices or because of lack of understanding/interest in the games. In our case, 37 percent of the risk-averse participants and around 29 percent of participants who are not risk-averse, invested the exact same amount in the probability, reward, and step-by-step games. We can rule out that the choice to invest the same amount is caused by the same position of the investment choice on the computer screen, as the order of the investment choices (0 CZK, 30 CZK, up to 150 CZK) was randomized in each game.} We also find that risk-averse individuals from wealthier backgrounds tend to invest the maximum in the probability game, but in the reward game, they tend to invest some interior positive amount, and that the risk-neutral or risk-loving individuals behave similarly in the probability and the reward games. 

To test for the difference in dispersion of choices between the two games, we use Levene's test on the equality of standard deviations. We can reject the hypothesis that the standard deviation of the decisions in the reward game is equal to or higher than the standard deviation of the decisions in the probability game (p=0.036). In other words, the decisions in the probability game are significantly more dispersed compared to the decisions in the reward game.

\begin{table}[H]
	\centering
	\begin{threeparttable}	
		\begin{tabular}{lccccccccc}
			             & \\
             &    Risk averse&     Not risk averse\\ \hline
  Invested 0 CZK&    \textbf{0.0295}&  -0.0109\\ & {(0.014)} & {(0.022)}  \\
 Invested 30 CZK&    -0.0127 &  -0.0437\\ & {(0.022)} & {(0.027)}  \\
 Invested 60 CZK&  \textbf{-0.0759}&   -0.0164\\ & {(0.028)} & {(0.031)} \\
 Invested 90 CZK&  \textbf{-0.0717}&   -0.0219\\ & {(0.032)} & {(0.036)} \\
Invested 120 CZK&  \textbf{0.0464}&    -0.0164& \\ & {(0.025)} & {(0.033)}  \\
Invested 150 CZK&  \textbf{0.0844}& \textbf{0.1093}& \\ & {(0.030)} & {(0.035)} \\
               n&                474&               366\\

		\end{tabular}

	    \begin{minipage}{\linewidth}\footnotesize 
			Note: {Invest 0 (up to 150) represents the dependent variable in a probit regression regressed on a treatment dummy, which equals 1 for the probability game and 0 for the return game. The estimated coefficients represent the marginal values and measure how much more likely people are to invest if they play the probability game compared to the reward game. The numbers in brackets are standard deviations. The columns show the results for risk averse/not risk averse people with full understanding.}
		\end{minipage}
	\end{threeparttable}
	\caption{Differences in investment probabilities between the two games}
	\label{tab1:probit_tab1}

\end{table}

Next, we run six probit regressions for each investment decision separately (see Table \ref{tab1:probit_tab1}). The dependent variable in each regression is a dummy variable equal to 1 if the investment was selected and 0 otherwise. Our independent variable is a dummy variable which equals 1 if the decision is made in the probability game, and 0 if the decision is made in the reward game. The estimated coefficients represent the marginal values and measure how much more likely people are to select the decision in the probability game relative to the reward game. Standard errors, clustered at the individual level, are displayed in brackets. Risk-averse individuals are significantly more likely to invest 0, 120 and 150 CZK in the probability game compared to the reward game (p=0.033, p=0.062 , and p=0.005, respectively), and are significantly more likely to invest 60 CZK or 90 CZK in the reward game compared to the probability game (p=0.006, and p=0.026 respectively). In the reward game, the participants tend to invest some positive amount, which is consistent with the interior solution predicted by the model, but in the probability game, the same individuals tend to invest closer to the corners. We do not see any patterns among non-risk-averse participants (shown in column 2 in Table \ref{tab1:probit_tab1})}. The results hold even for the participants with good (rather than full) understanding, if other risk aversion measures are used, and if we control for the order of the games and other covariates (such as gender, age, education, personal income, etc.). The results are presented in {Appendix Tables \ref{AppTab6:probit_understanding}, \ref{AppTab8:probit_covar}}, and \ref{AppTab7:probit_diffRAmeasures}.\footnote{People with poor or no understanding of the games show no shift in their investment decisions between the probability and return games. The results are available upon request.}

One may be concerned about the external validity of our experimental evidence. Specifically, individuals from the Czech Republic sample may be substantially better off than a typical poor person who is the target of microfinance. To alleviate this concern and to shed light on whether our theory and empirical evidence do capture the behavior of the very poor, we divide our sample into three groups by income: low, medium and high-income.\footnote{We use respondents' self-declared personal and household income. A low-income household is defined as having a monthly income of less than 20,000 CZK. For reference, households in the two poorest deciles earned on average  19,782 CZK in December 2020 and 21,082 CZK represented the monthly income poverty line for a household of 2 adults in the Czech Republic (Czech Statistical office). A medium-income household is defined by income between 20,001 and 40,000, and a high-income household by income above 40,000 CZK. For reference, the average household income is 38,020 CZK in December 2020.} We repeat the six regressions from Table \ref{tab1:probit_tab1} for each income group and present our results in 
Table \ref{AppTab9:probit_diffHHinc}. 

In the reward game, respondents at all income levels were significantly more likely to make an investment but not to invest the maximum possible. The picture is significantly different in the probability game: the three columns in the table indicate that the rich are more likely to invest the maximum amount in the probability game, more than in the reward game (Column 3), for the poor the opposite is true: they are more likely to invest less (typically zero) in the probability game compared to the reward game (Column 1). Participants from the middle-income group make significantly more investment decisions in both corners - minimum or maximum amounts - in the probability game in comparison to the reward game.  

Thus, if poor households in our sample, which are wealthier than the typical poor household in a poor country, do not invest, then much poorer households - such as those targeted by microfinance - would also avoid such an investment. In reaching this conclusion, we follow the conventional claim that the poor in poor countries are not fundamentally different than other people, they simply have less income. Thus, our theory and empirical evidence suggest that the poor in poor countries would choose not to borrow and invest if, as we propose, investment affects the probability of success.

\begin{table}[H]
	\centering
	\begin{threeparttable}	
		\resizebox{\textwidth}{!}{
		\begin{tabular}{lccccccccc}
			             &    Low HH income  &     Medium HH income &       High HH income  \\ \hline\hline
Invested 0 CZK&    \textbf{0.0741}&       \textbf{0.0303}&       -0.0351 \\ & {(0.029)} & {(0.017)} & {(0.025)} \\
Invested 30 CZK&    0.0494    &   \textbf{-0.0606}&     -0.0175 \\ & {(0.043)} & {(0.028)} & {(0.047)}  \\
Invested 60 CZK&  \textbf{-0.0741}&       \textbf{-0.0909}&   -0.0526\\ & {(0.042)} & {(0.046)} & {(0.058)} \\
Invested 90 CZK&   \textbf{-0.0864}&        0.0101&    \textbf{-0.1930} \\ & {(0.047)} & {(0.049)} & {(0.077)} \\
Invested 120 CZK&    0.0617&      <-0.0001&   \textbf{0.1053} \\ & {(0.041)} & {(0.035)} & {(0.060)} \\
Invested 150 CZK&   -0.0247&   \textbf{0.1111}&      \textbf{0.1930}\\ & {(0.046)} & {(0.047)} & {(0.064)} \\
               n&                162&               198&          114 \\ \hline

		\end{tabular}
		}
		%\begin{minipage}{\linewidth}\footnotesize \footnotesize
			%Note:
		%\end{minipage}
	\end{threeparttable}

        \begin{minipage}{\linewidth}\footnotesize 
      
       Note: The estimated coefficients represent the marginal values and measure how much more likely are people to invest in the amount if they play probability game compared to the return game based on their household income. The numbers in brackets are standard errors. 
    \end{minipage}
	\caption{Differences in investment probabilities between the two games, by household income}
	\label{AppTab9:probit_diffHHinc}
\end{table}

We use the step-by-step game to test our third prediction: participants prefer to invest in increasing the probability of success over investment in increasing the reward if successful, for the same expected return. As described above, the participants make four consecutive decisions, in which they either increase the probability of winning a fixed reward, or increase the reward, while keeping the probability unchanged. The decision paths in the step-by-step game allow us to check whether the second-order stochastic dominance prediction holds. We pool the four decisions and see that 70.4\% of the risk-averse participants invested predominantly in the probability of winning (they increased the probability three to four times) and the percentage is a bit higher (72.3\%) for the very risk averse. Among others, 9.9\% of the sample invested in the reward game and the remaining 19.8\% split their investment equally into return and probability. We conclude that if people have an option to either increase the probability of success or increase their reward, they strongly prefer the first option.

\section{Conclusion}

High hopes of significantly reducing poverty through microfinance were dashed by disappointing results. Our main contribution is in proposing a theory that could explain why the poor tend to turn down the opportunity to borrow money to pursue high return investments. The theory is based on risk aversion, which is central in the life of the poor, and does not rely on fixed costs or any other non-convexity in the production technology. Non-convexities are commonly used in the related theoretical literature, but seem to be inconsistent with the facts. Our theory is consistent with existing evidence and supported by the findings from our experiment.

We believe that for effective poverty-reduction policy, identifying the effect of microcredit on business creation and on reduction in poverty is crucial but insufficient. Understanding the reason for its failure is important, in particular because the evidence is inconsistent with the existing theoretical literature. We hope that our theory and findings bring us a step closer to understanding the issue, provide guidance to future empirical studies and identify useful policy implications. 

The main policy implication that we can draw from our theory, and that should be further tested, is that to facilitate investment by the poor, policy should be aimed at reducing the risk they face. One example would be to condition repayment of debt on outcomes, with a higher interest rate when investment yields successful outcomes, and forgiveness of most of the debt in case of failure. Consistent with this,
\cite{battaglia2018repayment}  find that repayment flexibility improves businesses outcomes via risk taking. It is worth noting that in our experiment women invest less than men, consistent with existing evidence that women are more risk averse. Therefore, special attention could be given to reducing the risk attached to investments made by women.
%Battaglia, Gulesci and Madestam (2018)

Another reasonable conclusion is that the path towards economic growth and reduction of poverty isn't more small businesses but rather, as in developed countries, more jobs that pay higher wages.  Perhaps, as proposed by \cite{banerjee2015miracle} the majority of the poor lack the training and skills to be successful entrepreneurs and business owners, and there is less potential for high return businesses run by the poor than is anticipated by microcredit enthusiasts. However, it could be the case that the major hurdle to investment in businesses is the high level of risk aversion among those who live in extreme poverty. In this case, offering credit to established businesses could be a promising approach, as these businesses could grow and create more jobs.\footnote{A positive effect of microcredit on established business was identified by several studies, e.g., \cite {meager2022aggregating}.} Perhaps, therefore, microcredit does provide a significant benefit of an increase in labor demand and wages, but this is of course beyond the focus of our paper.\footnote{This benefit, even if significant, cannot be identified by most empirical studies, because the randomized control trials designed to test the effect of microcredit cannot identify general equilibrium effects. \cite{breza2021measuring} investigated what happened in Andhra Pradesh, India, when microcredit institutions were shut down in 2010. They found that this was followed by a notable decrease in wages in rural areas.}
  
\newpage 
%\newpage
\bibliographystyle{abbrvnat}
 % \bibliographystyle{jmb}
%\bibliography{_KhazanovMoavNeemanZoabi17}
\bibliography{_CKMNZ22}

\newpage 

\appendix
\appendixpage
\addappheadtotoc
 
\renewcommand{\theequation}{\Alph{section}.\arabic{equation}}
\setcounter{equation}{0}

\renewcommand{\thefigure}{\Alph{section}.\arabic{figure}}
\setcounter{figure}{0}

\renewcommand{\thetable}{\Alph{section}.\arabic{table}}
\setcounter{table}{0}

\section{Proofs\label{Appendix}}

\noindent {\bf Proof of Proposition 1}

\bigskip

\bigskip

\noindent \textbf{Proof.} We show that if $U$ is increasing at some
$p$, then it is increasing for all $p'>p$. The agent's expected
utility is equal to
\[
U(p)\equiv-pe^{-\lambda(H-\alpha p)}-(1-p)e^{-\lambda(L-\alpha p)}
\]
and its derivative with respect to $p$ is equal to
\[
U'(p) = \left(1+\lambda\alpha p\right)e^{\lambda\alpha p}\left(e^{-\lambda L}-e^{-\lambda H}\right)-\lambda\alpha e^{\lambda\alpha p}e^{-\lambda L}.
\]

\noindent Suppose that $U$ is increasing at $p$, or $U'(p)>0$. It follows
that
\[
\begin{array}{ccl}
U''(p) & = & \lambda\alpha\left[\left(2+\lambda\alpha p\right)e^{\lambda\alpha p}\left(e^{-\lambda L}-e^{-\lambda H}\right)-\lambda\alpha e^{\lambda\alpha p}e^{-\lambda L}\right]\\
 & = & \lambda\alpha\left[U'(p)+e^{\lambda\alpha p}\left(e^{-\lambda L}-e^{-\lambda H}\right)\right]\\
 & > & 0,
\end{array}
\]
which implies that $U'(p')>U'(p)$ for all $p'>p$ or that $U$ continues
to increase throughout the remainder of its range.\hfill{}$\blacksquare$

\bigskip\bigskip

\noindent {\bf Proof of Proposition 2} 

\bigskip 

%\noindent \textbf{Proposition 2.} \emph{The agent's choice of $p$ is
%discontinuous in its level of risk aversion $\lambda$. There exists a threshold level of risk aversion $\lambda^{o}>0$ such that more risk
%averse agents with $\lambda>\lambda^{o}$ choose the minimal probability
%$p=0$, and less risk averse agents with $\lambda<\lambda^{o}$
%choose the maximal probability $p=\overline{p}$. }

\bigskip
 \noindent \textbf{Proof.}  Recall that von Neumann-Morgenstern utility functions are only unique up to affine transformations. Hence, the utility function $u(x)=-e^{-\lambda x}$ is equivalent to the utility function $u(x)=\frac{1-e^{-\lambda x}}{\lambda}$, which converges to a linear function as $\lambda$ tends to zero by L'H\^{o}pital's Rule. This implies that the agent becomes risk neutral and so prefers the lottery $(H-\alpha\overline{p},L-\alpha\overline{p};\overline{p},1-\overline{p})$ over the certain outcome $L$. By continuity, this is also the case for all agents with small enough level of risk aversion $\lambda$.

An agent prefers the certain outcome $L$ over the lottery $(H-\alpha\overline{p},L-\alpha\overline{p};\overline{p},1-\overline{p})$ if and only if
\[
-pe^{-\lambda(H-\alpha p)}-(1-p)e^{-\lambda(L-\alpha p)} < -e^{-\lambda L}
\]
if and only if
\[
pe^{\lambda(\alpha p +L -H)}+(1-p)e^{\lambda\alpha p} > 1.
\]
As $\lambda$ increases to infinity, $e^{\lambda(\alpha p +L -H)}$ tends to zero, but $ e^{\lambda\alpha p}$ tends to infinity, which implies that the last inequality is satisfied for all $\lambda$ large enough.

Finally, the fact that an agent with a smaller $\lambda$ is less risk averse than an agent with a larger $\lambda$ implies that any lottery that is preferred over a certain outcome by the former is also preferred by the latter. It therefore follows that there exists a threshold level of risk aversion $\lambda^{o}>0$ such that more risk
averse agents with $\lambda>\lambda^{o}$ choose the minimal probability
$p=0$, and less risk averse agents with $\lambda<\lambda^{o}$
choose the maximal probability $p=\overline{p}$.\hfill{}$\blacksquare$

\bigskip\bigskip

\noindent {\bf Proof of Proposition 3} 

\bigskip 
 
\bigskip

\noindent \textbf{Proof.} The first and second derivatives of \emph{$V(c)$}
are given by\emph{ }
\[
V'(c)=pu'(B+H(c)-c)\left(H'(c)-1\right)-(1-p)u'(B+L-c)
\]

\noindent and
\[
V''(c)=pu''(B+H(c)-c)\left(H'(c)-1\right)^{2}+pu'(B+H(c)-c)H''(c)+(1-p)u''(B+L-c),
\]
respectively. The conclusion follows from the concavity of the functions
$u\left(\cdot\right)$ and $H\left(\cdot\right)$.\hfill{}$\blacksquare$

\bigskip\bigskip

\bigskip\bigskip

\noindent {\bf Proof of Proposition 4} 

\bigskip

\bigskip

\noindent \textbf{Proof.} Follows immediately from the fact that the $p$-lottery second-order-stochastically-dominates the $p'$-lottery.\hfill{}$\blacksquare$

\bigskip\bigskip

\section{CRRA utility and a simple calibration} \label{Appendix_qunuantitative}

In this section we show that the discontinuity result may also hold for a CRRA utility function. However, as the numerical example below illustrates, the local maxima can be interior rather than corner, and may change with the initial wealth. Specifically, richer people may make higher investments. Then we provide a simple calibration, using CRRA utility function and data from \cite{AugsburgDeHaasHarmgartMeghir15}, which suggests that, under reasonable values of relative risk aversion, poor agents do not invest.  

\subsection{The case of CRRA}\label{CRRA}

Assume a CRRA utility function of the form $u(x)=\frac{x^{1-\sigma}-1}{1-\sigma}$. In this case, our expected utility function that corresponds to the one in section \ref{sec:section2} becomes\footnote{We denote by $\mu$ the cost of investment and keep, in this subsection, the assumption made in our model, $\mu=c(p)=\alpha p$.}

\begin{align} \label{Eq_CRRA}
U(p)=p \frac{\left(B + H - \mu\right)^{1-\sigma}-1}{1-\sigma} + (1-p) \frac{\left(B + L - \mu \right)^{1-\sigma}-1}{1-\sigma}
\end{align}

\begin{figure}[H]
\begin{center}
\includegraphics[scale=0.35]{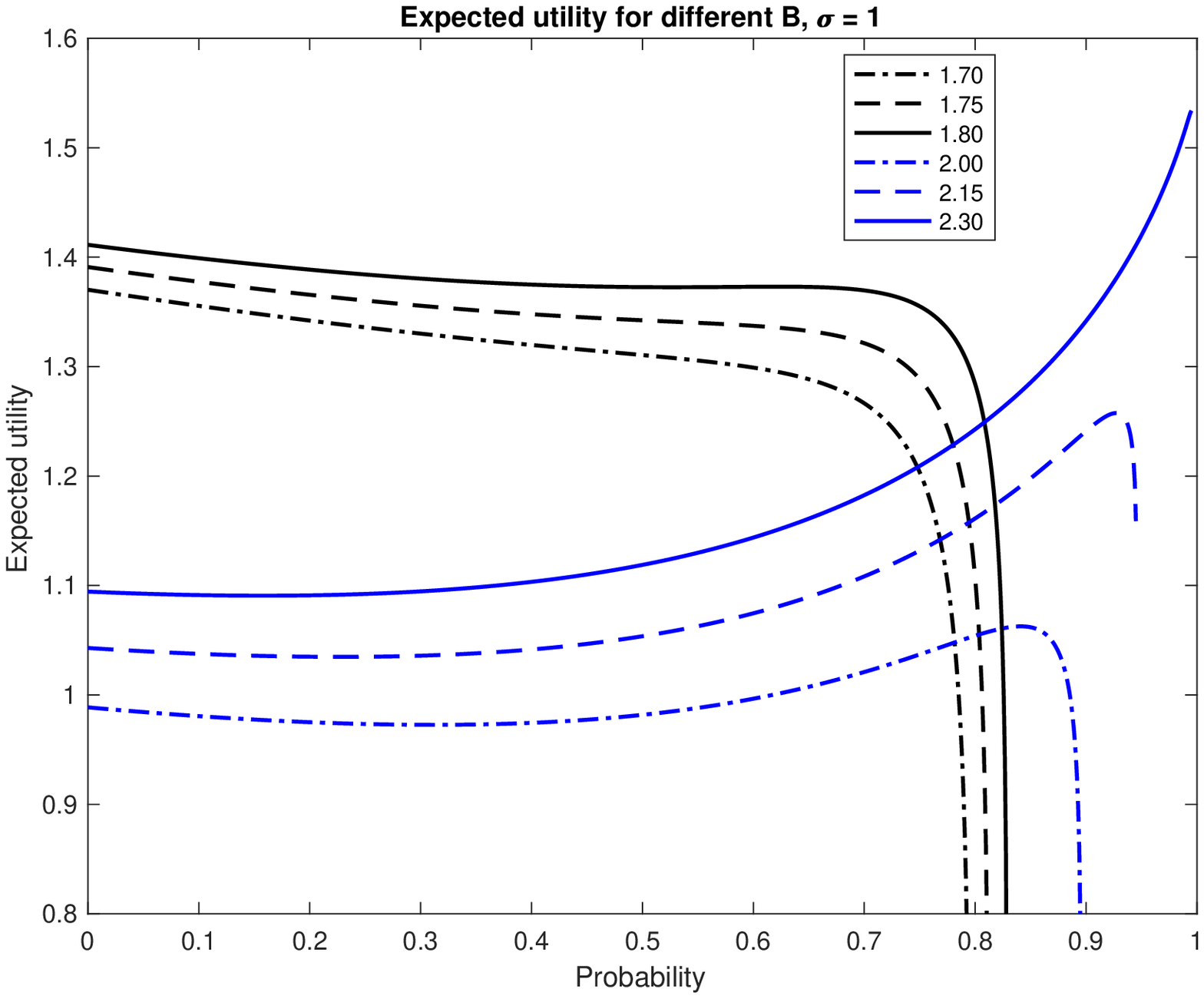}
\includegraphics[scale=0.35]{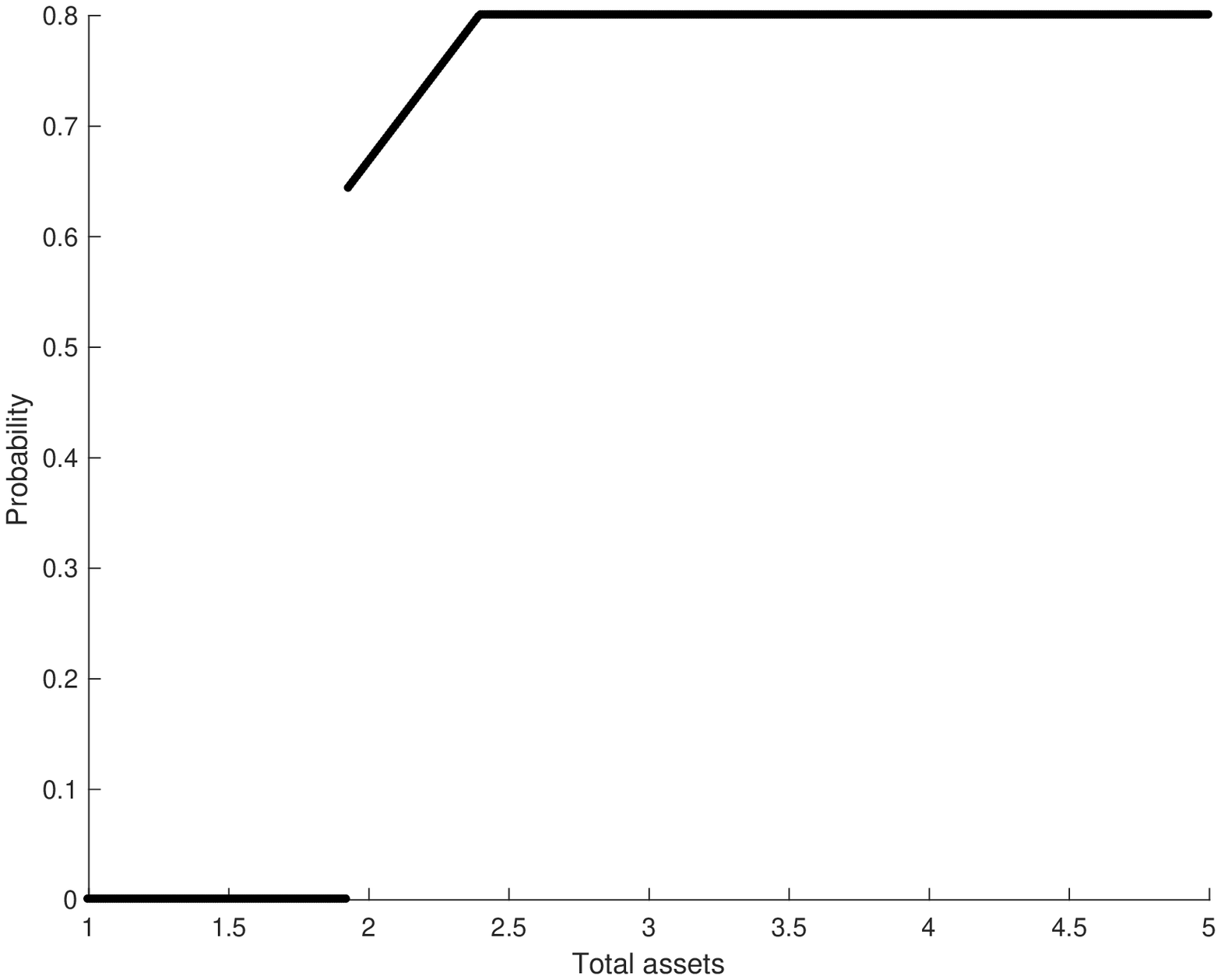}
\end{center}
\caption{\label{Fig_CRRA} Left panel: Expected utility as a function of probability (levels adjusted by constant for convenience of presentation on one figure). Right panel: Optimal probability.}
\end{figure}

Figure \ref{Fig_CRRA} draws the expected utility for different levels of wealth , which corresponds to the values $\sigma = 1$ , $H = 5.5, L = 0.6875, \alpha= 3$, $\bar{p} = 0.8$ .\footnote{This corresponds to $u(x)=\ln(x)$.}

The left panel of Figure \ref{Fig_CRRA} shows that while for a relatively low initial wealth agents (black curves) the optimum is achieved at $p=0$, for a relatively high initial wealth agents (blue curves) the optimum is achieved at strictly positive and relatively high level of $P$. More importantly, the U-shaped pattern still exists but it does not have to occur allover the range $[0,1]$ (blue curves). The right panel of Figure \ref{Fig_CRRA} shows that the discontinuity in optimal choice of probability still holds. Specifically, for a relatively low initial wealth, $B<1.93$ agents choose not to invest. Finally, throughout the range $1.93<B<2.40$, the optimal value of $p$ increases with wealth.\footnote{Remember that in the CARA case the U-shaped pattern holds for the whole range between 0 and 1. This implies that if it is optimal for the agent to invest, it occurs at $\bar{p}$. This made our analytical problem tractable.}

\subsection{A Simple Calibration}\label{Sec_SimpleCalibration}

We use data from \cite{AugsburgDeHaasHarmgartMeghir15} who studied the effect of microcredits in Bosnia. They conducted an experiment by providing randomly loans to those who were rejected by MFI. To facilitate their study, the authors collected data on various socioeconomic variables, ranging from household consumption and assets to income and savings choices. 

The data we use includes: 1) \textbf{assetvalue} which corresponds to Endowment $B$ in the model;  2)  \textbf{y\_max} which corresponds to the return H in the good state of the world; 3) \textbf{bm\_expenses} which corresponds to investment $\mu$. ; 4) \textbf{past\_success} which is used to compute probability of success P.\footnote{
Variable corresponding to $B$ was calculated by the authors as the total value of assets owned by the respondents. $H$ comes from the question  ``\textit{Imagine that you do receive the loan from EKI and have
	a very good month/year, economic conditions are flourishing and stable
	and there is great demand for your product/service\dots{} What would
	be the }\textbf{\textit{maximum}}\textit{ amount of profit this business
	of yours receives in such a situation over the next month/year?}''. $\mu$ corresponds to the question ``\textit{average yearly expenses of main business}''
	$P$ is calculated based on the question ``\textit{Please respond to the following statements on a scale of 1 (Disagree) 2 (Neutral) 3 (Strongly Agree).	Previous year was successful financially}''.  
	}

To calculate the probability of success, $p$, we use the question \emph{previous year was successful financially}. We assume that those who strongly agreed with the statement could be considered successful. Then, we create a dummy variable that takes value 1 if a person was successful and 0 -- otherwise. Next, we sort the sample based on the asset value $B$, from smallest to the largest. Then, we divide this sorted sample into 30 income groups by the value of total assets that they own: the first group contains 18 poorest respondents, the second -- the next 18 poorest poorest, etc. Finally, for each group we compute the share of people within a group who were successful. Such share is an estimate of the  probability of success $p$ for a given income group. We notice that the minimal value of $p$ is $0.(22)$, which corresponds to the lowest income group and the maximal value is $0.88$, which corresponds to the highest income group. Similarly, we aggregate $\mu$ and $B$, by taking the mean of individual $\mu$ and $B$, respectively, for a given income group.

For the purpose of the calibration we assume that the probability is a linear function of investment $p = \gamma \mu$, $L = 0$ and take $\bar{p}=0.88$, $H=62 \times 10^3$ from the data. We also assume that the maximum $\mu$ from the data corresponds to $\bar{p}$ and use it to calculate $\gamma$.

We plot the probability of success as a function of total assets for different levels of $\sigma$ in the range considered acceptable in the literature. The figure illustrates our main results: for reasonably low levels risk aversion poor individuals choose to avoid investment, and a continuous rise in wealth leads to a discontinuous jump in investment.

\begin{center}
	\begin{figure}[H]
		\centering
		\includegraphics[scale=0.8]{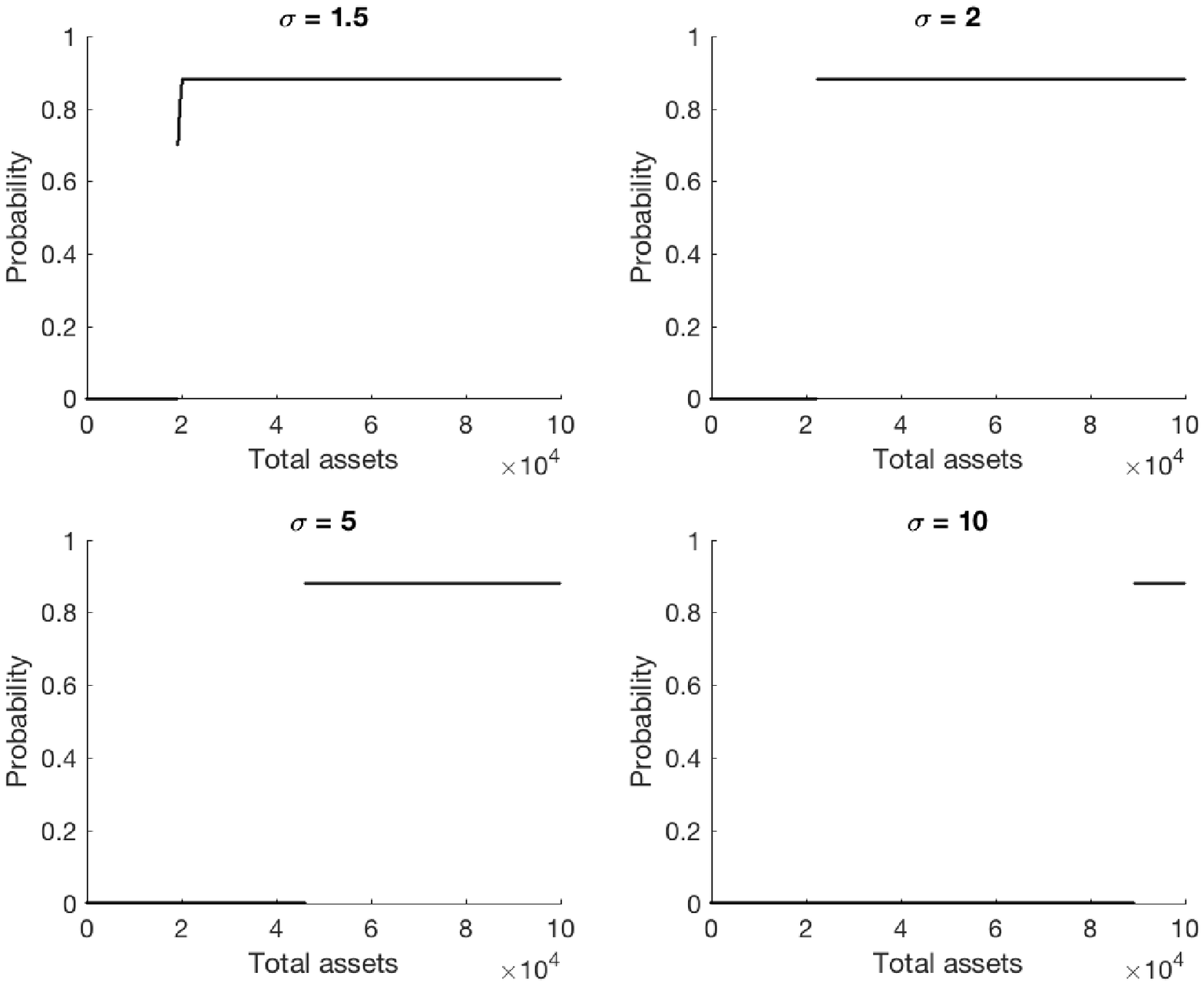}
		\caption{$H=62 000$, $L=0$, $\underline{p} = 0$, $\bar{p} = 0.88$ }
		\label{fig:newtake_fig1_v2}
	\end{figure}
\end{center}

\newpage 
\clearpage
\pagenumbering{arabic}

\section*{Online Appendix}

\section{A Model with a Resource Constrained Agent [For Online Publication]}  \label{App_Resource_Constrain}
\setcounter{equation}{0}
\setcounter{figure}{0}
 \setcounter{table}{0}
 
In this section, we show that the results of the simple model continue to hold when the model is embedded in a competitive financial market with the possibility of bankruptcy (when the investment project fails and the agent pays only part of the debt). As in the simple model, individuals can invest in the probability of success of a risky project with an exogenous binary outcome of high or low income. They can augment their investment with a loan and pay the competitive risk-adjusted interest rate in a perfect loan market, which takes into account the probability of a low-income outcome and bankruptcy. We assume that individuals' wealth is unobserved by the financial intermediary, and the interest rate is determined by a zero-profit condition, under limited liability: only the income from the investment project can be used to pay back the debt.

We show that an individual's choice of investment is discontinuous in its degree of risk aversion, and therefore in wealth, assuming that risk aversion declines with wealth. This implies that individuals face a tradeoff between a safe, low return option, and a risky, high expected return option. If an individual chooses the safe option, her initial wealth is augmented by low income. If she invests in the project, the end outcome in case of failure is that she is left with her initial wealth and the low income from the unsuccessful project, net of the investment cost. If this initial wealth is low then risk aversion is high: the disutility from losing the low income in the risk free option is high. This leads to our main result: despite the absence of non-convexities in the production technology, optimization implies that the poor behave as if there is a fixed cost which prevents them from investing in a high return project despite the fact that credit is available at a competitive rate.

Suppose that the agent has an initial income of $B\geq0$ that it can use in order invest in a risky binary project as described in the previous section. An investment of size $c(p)=\alpha p$ generates an additional income $H$ with probability $p$ and an additional income $L$, $0 < L<H$, with probability $1-p$, where $0<\alpha<H-L$.

An investment that is larger than $B$ requires the agent
to borrow. Suppose that the agent has access to a competitive
credit market in which the riskless interest rate is normalized to
zero. A loan of size $b\geq0$ can be obtained at the interest rate
$r(b)$ that allows lenders to break even. We assume that $B$ is
non-verifiable to lenders so that an individual who borrows any amount
return a maximum amount $L$ if its additional income is realized to be $L$,
and a maximum amount $H$ if it is realized to be $H$. We assume that the success of the project as well as the agent's choice of probability $p$ are verifiable
(for example, because the project requires investment in observable physical capital) so
that lenders are able to asses the correct interest rate to charge
the loan, and would refuse loans that are larger than what is needed
in order to finance the agent's investment.\footnote{\label{footnote9}Note that the agent may want to borrow a larger amount than the amount necessary to finance its investment because such a loan provides insurance to the agent: an agent who borrows such a larger amount enjoys a certain income that is paid back only upon success. Lenders may be reluctant to lend larger sums because of moral hazard considerations, and in any case, the point of this paper is that poor agents cannot reduce or eliminate their exposure to risk, which such larger loans would facilitate.} Finally, we assume that in case of indifference, the agent prefers a larger to a smaller loan (this last assumption is not necessary for our results, but it simplifies the analysis below).

Suppose that the individual has a CARA utility function and chooses
the probability $p$ at cost $c(p)=\alpha p$ as described above.

\bigskip

\noindent \textbf{Proposition 5.} \emph{The agent finances its entire investment $c(p)=\alpha p$ through a loan.}

\bigskip

\noindent \textbf{Proof.} If $c(p)\leq L$ then the individual is indifferent with respect to how it finances
its investment because regardless of whether it finances the investment from its own funds or through a loan,
its income is $B+H-c(p)$ and $B+L-c(p)$
following success and failure of the project, respectively. It therefore follows
that if $c(p)\leq L$ then the individual
will finance its investment entirely through a loan.

If $c(p)>L$ then borrowing allows the individual to insure itself
against risk because an individual who borrows the entire amount necessary
for investment $c(p)$ returns only $L$ if the project fails and so enjoys an income of $B$ in that case, whereas an individual
who borrows a smaller amount and relies on its own funds (but still
chooses the same probability $p$) still has to pay back $L$
if the project fails so only enjoys a smaller income
than $B$ in that case (note that a same probability $p$ produces equal expected incomes in the two cases and that lenders earn zero profits). A bigger loan implies that the induced lottery second-order-stochastically-dominates the lottery induced by a smaller loan. So every risk averse individual would prefer a bigger loan over a smaller loan. It therefore follows that in this case the agent would borrow the largest amount possible, which is equal to $c(p)$.\footnote{The reasoning above implies that the agent would like to borrow possibly even more than $c(p)$, but we assume that lenders would refuse larger loans (see the discussion in footnote \ref{footnote9}).}$^{,}$\footnote{Lenders will obviously not lend more than $\min\{c(p),(1-p)L+pH\}$. However, the assumption that $(1-p)L+pH-c(p) > L$ ensures that this minimum is obtained on $c(p)=\alpha p$.}\hfill{}$\blacksquare$

\bigskip

We show that, as in the case described in the simple model, the agent's expected utility is U-shaped in $p$. The fact that individuals borrow the entire amount needed to finance their investment implies that the individual's induced expected utility
function $U(p)$ is given by:
\begin{equation}
U(p)=\left\{ \begin{array}{lcl}
pu(B+H-\alpha p)+(1-p)u(B+L-\alpha p) &  & \textrm{if }p<\frac{L}{\alpha}\\
pu(B+H-L-\alpha+\frac{L}{p})+(1-p)u(B) &  & \textrm{if }\frac{L}{\alpha}\leq p
\end{array}\right.\label{eq:U(p)}
\end{equation}
because an individual who borrows an amount $c(p)=\alpha p < L$ returns $\alpha p$ in both states of the world, and an individual who borrows an amount $c(p)=\alpha p \geq L$ returns $L$ if the project fails, and $L+\alpha-\frac{L}{p}$ if the project succeeds, so that $p\left(\frac{\alpha p -(1-p)L}{p}\right)+(1-p)L=\alpha p$
overall.\footnote{Observe that $\frac{\alpha p -(1-p)L}{p}\leq\frac{pH+(1-p)L-(1-p)L}{p}=H$
so the individual can indeed return the loan if the project succeeds.}

\bigskip

\noindent \textbf{Proposition 6.} \emph{The individual's induced expected utility function $U(p)$ that is described in (\ref{eq:U(p)}) is U-shaped
in $p$.}

\bigskip

\noindent \textbf{Proof.} For values of $p$ that are such that $c(p)<L$
or $p<\frac{L}{\alpha}$, it is possible to show that the function
$U(p)=pu(B+H-\alpha p)+(1-p)u(B+L-\alpha p)$ is U-shaped using a
similar argument to the one used in the proof of Proposition 1.

For values of $p$ that are such that $c(p)\geq L$ or $p\geq\frac{L}{\alpha}$,
we need to show that the function $U(p)=pu(B+H-L-\alpha+\frac{L}{p})+(1-p)u(B)$
is increasing in $p$. This implies that $U(p)$ is
U-shaped over its entire range because $U(p)$ is continuous so the
argument is valid regardless whether $U(p)$ is decreasing, increasing,
or U-shaped for $p\in[0,\frac{L}{\alpha}]$.

The derivative of $U(p)$ with respect to $p\geq\frac{L}{\alpha}$
is equal to

\[
U'(p)  =  - pu'\left(B+H-L-\alpha+\frac{L}{p}\right)\frac{L}{p^{2}}+u\left(B+H-L-\alpha+\frac{L}{p}\right)-u(B).
\]

For our CARA utility function, $u(x)=-e^{-\lambda x}$ and $u'(x)=\lambda e^{-\lambda x}$. So, $U'(p)>0$ if and only if:

$$\lambda\left(H-L-\alpha +\frac{L}{p}\right)>\ln\left(1+\frac{\lambda L}{p}\right).$$
The conclusion follows from our assumption that $H-L > \alpha$ together with the fact that $x > \ln(1+x)$ for all $x>0$. \hfill{}$\blacksquare$

\bigskip\bigskip

\bigskip

It follows that like in the simple model, also in this model we have:

\bigskip

\noindent \textbf{Proposition 7.} \emph{The agent's choice of $p$ is
discontinuous in its level of risk aversion $\lambda$. There exists
a threshold level of risk aversion $\lambda^{o}$ such that more risk
averse agents with $\lambda>\lambda^{o}$ choose the minimal probability
$p=\underline{p}$, and less risk averse agents with $\lambda<\lambda^{o}$
choose the maximal probability $p=\overline{p}$.}

\bigskip

\noindent \textbf{Proof.} Follows from the same argument as in the proof of Proposition 2.\hfill{}$\blacksquare$

\bigskip

If agents with a smaller initial income (bequest) $B$ are also more
risk averse in the sense that their CARA utility function has a larger
risk coefficient parameter $\lambda$ then we have:

\bigskip

\noindent \textbf{Corollary.} \emph{The agent's choice of $p$ is
discontinuous in its initial income (bequest) $B$. There exists a threshold level
of income $B^{o}$ such that poorer agents who have a smaller initial income
$B<B^{o}$ choose the minimal probability $p=\underline{p}$, and
richer agents who have a larger initial income $B>B^{o}$ choose the maximal
probability $p=\overline{p}$.}

\section{\label{sec:appexp} Experiment [For Online Publication] }
\setcounter{equation}{0}
\setcounter{figure}{0}
 \setcounter{table}{0}
In this section, we provide additional details about the experimental design, and offer further supportive results. In subsection \ref{Sample} we (visually) describe four incentivized decisions of our respondents, provide randomization balance, and describe respondents' characteristics. In subsection \ref{Robust} we replicate the main results for various subsamples and we look at the distribution of the investment choices by gender. 

\subsection{Design, sample descriptive statistics, and randomization balance}\label{Sample}

\begin{table}[H] 
\begin{center}
    \includegraphics[scale=0.2]{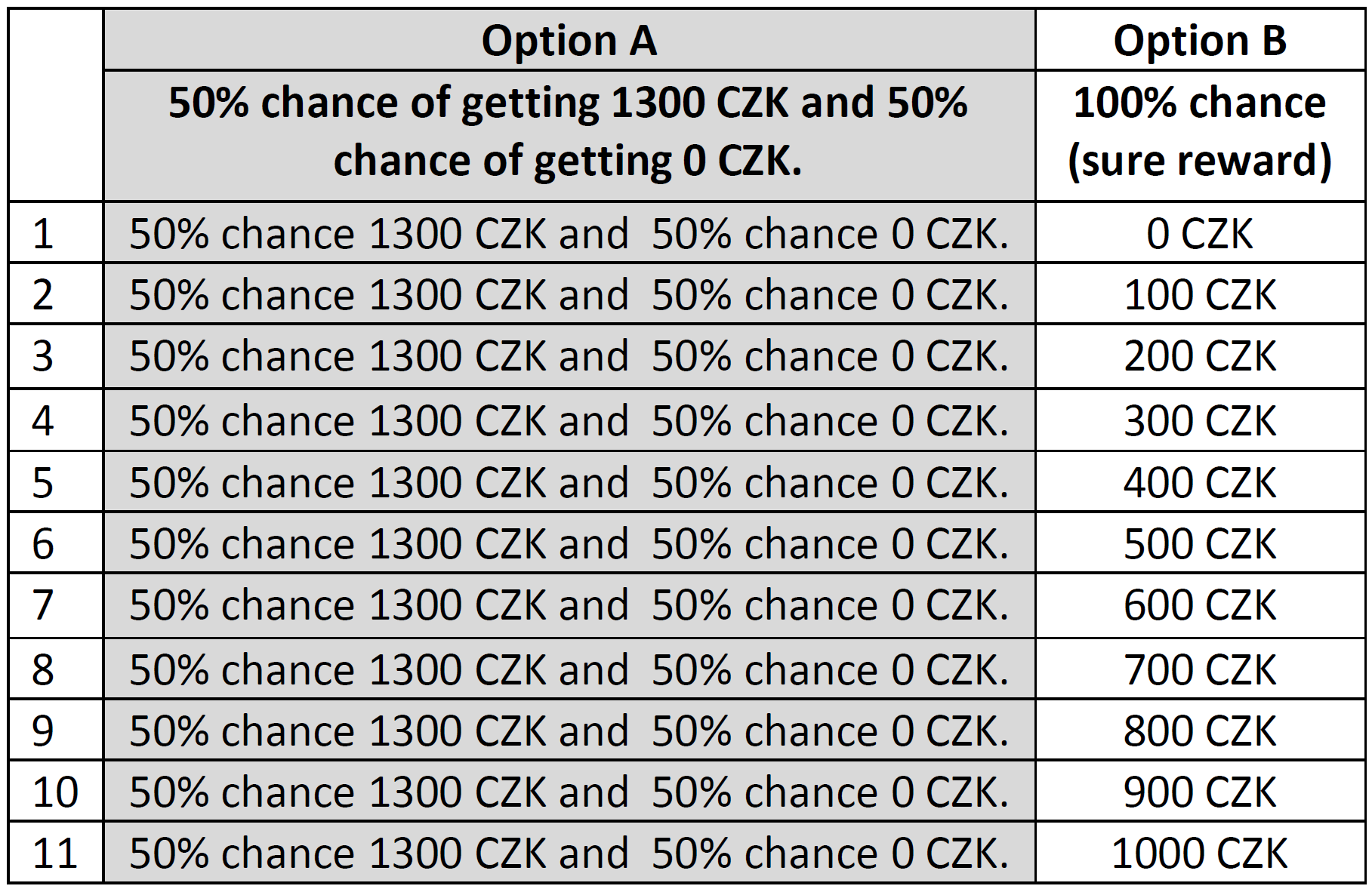}
    \caption{Decision 1: Risk elicitation}
	\label{AppTab1:RiskElicitation}
\end{center}

\end{table}

\begin{figure}[H] 

\begin{center}
    \includegraphics[scale=0.35]{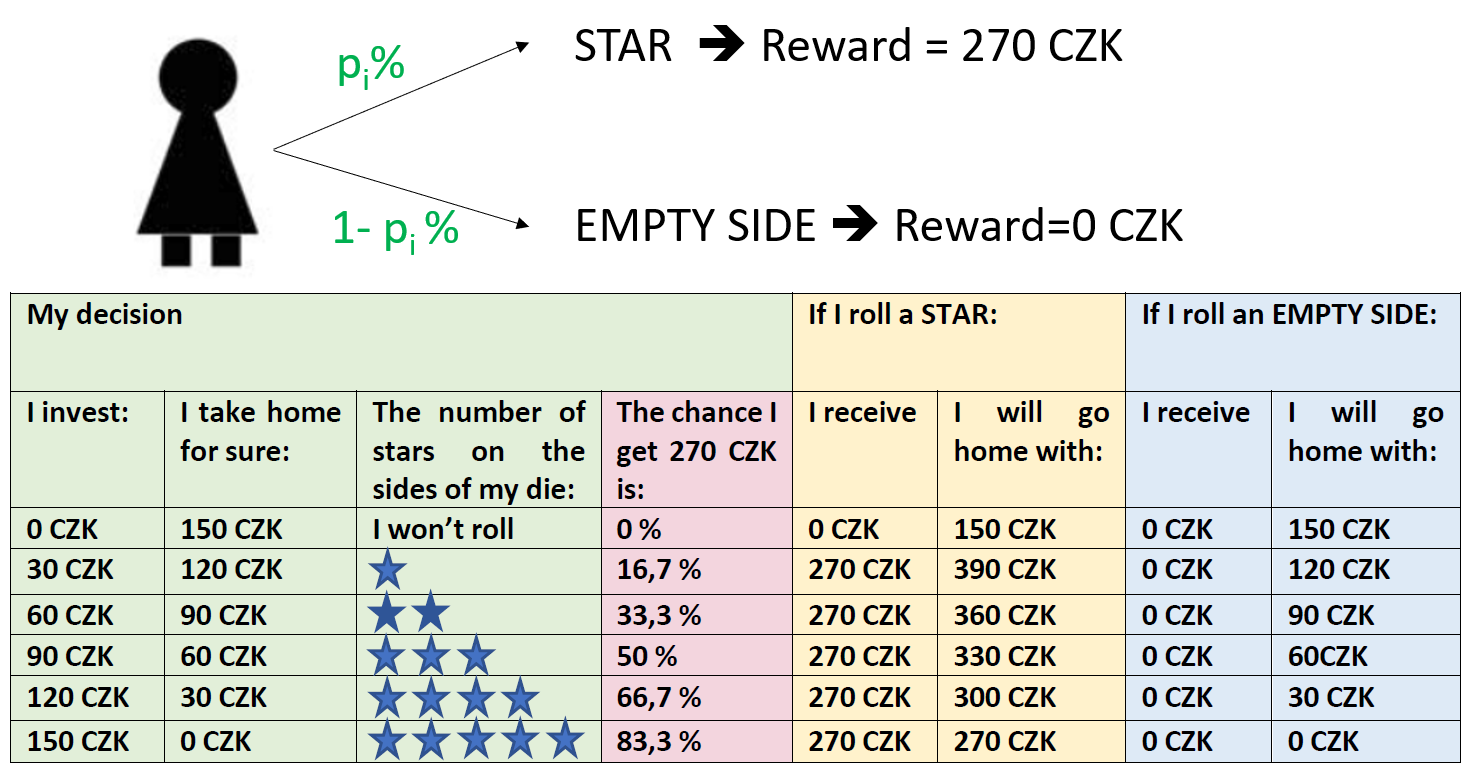}
    \caption{Decision 2/3: Investment decision in the Probability game}
	\label{AppFig1:ProbGameTable}
\end{center}
\end{figure}

\begin{figure}[H] 

\begin{center}
    \includegraphics[scale=0.29]{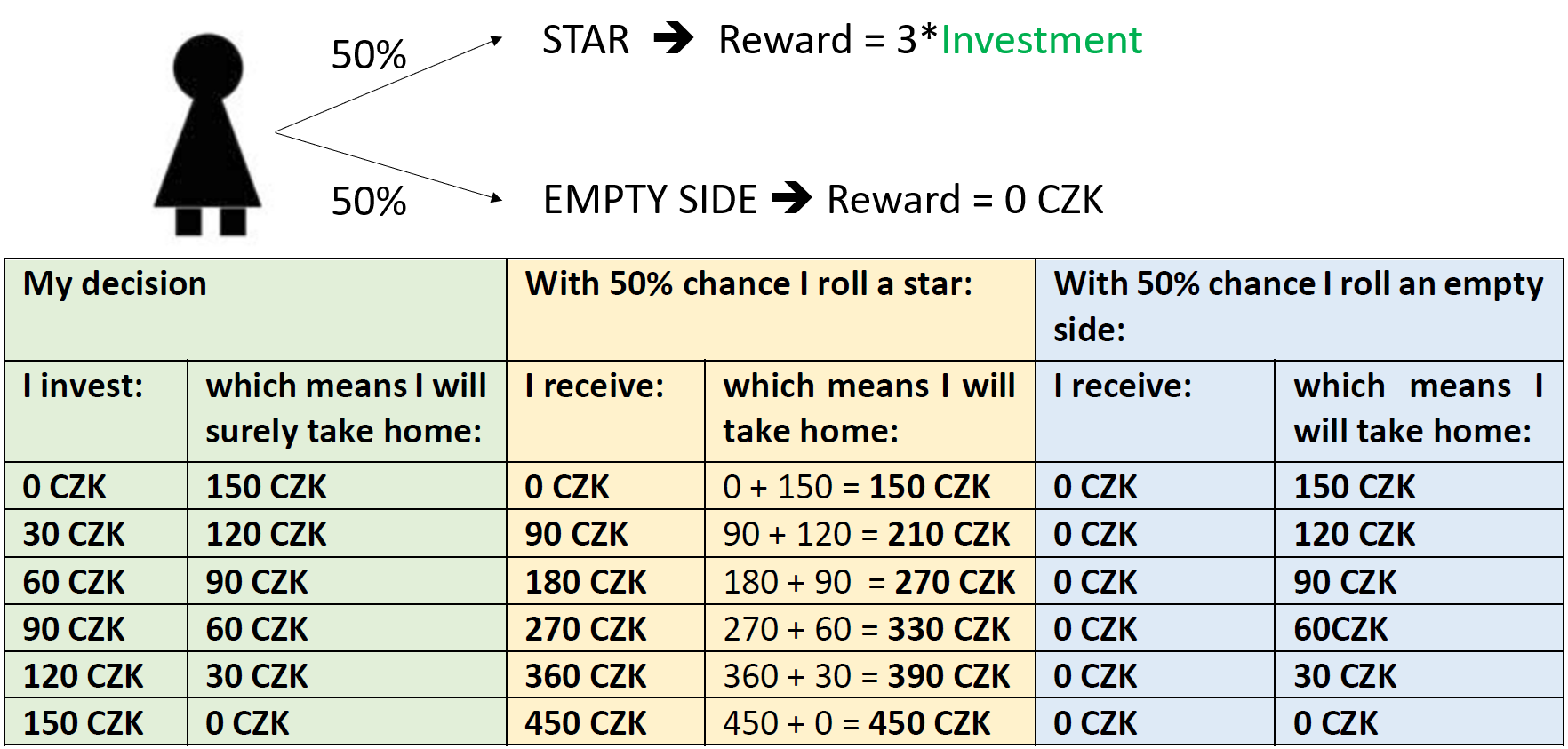}
    \caption{Decision 2/3: Investment decision in the Reward game}
    \label{AppFig2:RewGameTable}
\end{center}
\end{figure}

First decision of all participants was to decide from which raw in Table D.1 they prefered Option B over Option A. The order of the second and third incentivized decisions was randomized to rule out order effects. In each investment decisions the participants were endowed with 150 CZK which they could either kept for themselves or invest (a portion of) it. The participants received detailed description of the task followed by an example and a set of four control questions. The visual guides  in Figures \ref{AppFig1:ProbGameTable} and \ref{AppFig2:RewGameTable} were displayed on the screens in each decision step.

\begin{table}[H] 
\begin{center}
    \includegraphics[scale=2.25]{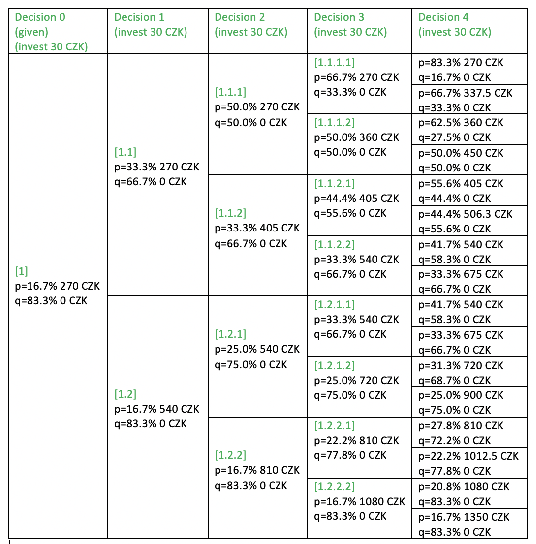}
    \caption{Decision 4: Investment decisions in the Step-by-step game}
    \label{AppTab2:StepByStep}
\end{center}
\end{table}

Table \ref{AppTab2:StepByStep} summarizes the portfolio of decision paths in the Step-by-step game. In Table \ref{AppTab3:DescrStat}, we provide descriptive statistics of the participants randomly selected into playing the reward game first (column 1) compared to those playing the probability game first (column 2). In column 3, we display mean differences between the two values, and the p-value as a result of the test testing whether the mean difference equals zero. Except for two variables (the number of people in the household, and full understanding of the games), there is no significant difference between the two groups.   

\begin{table}[H] 
\begin{center}
    \includegraphics[scale=0.75]{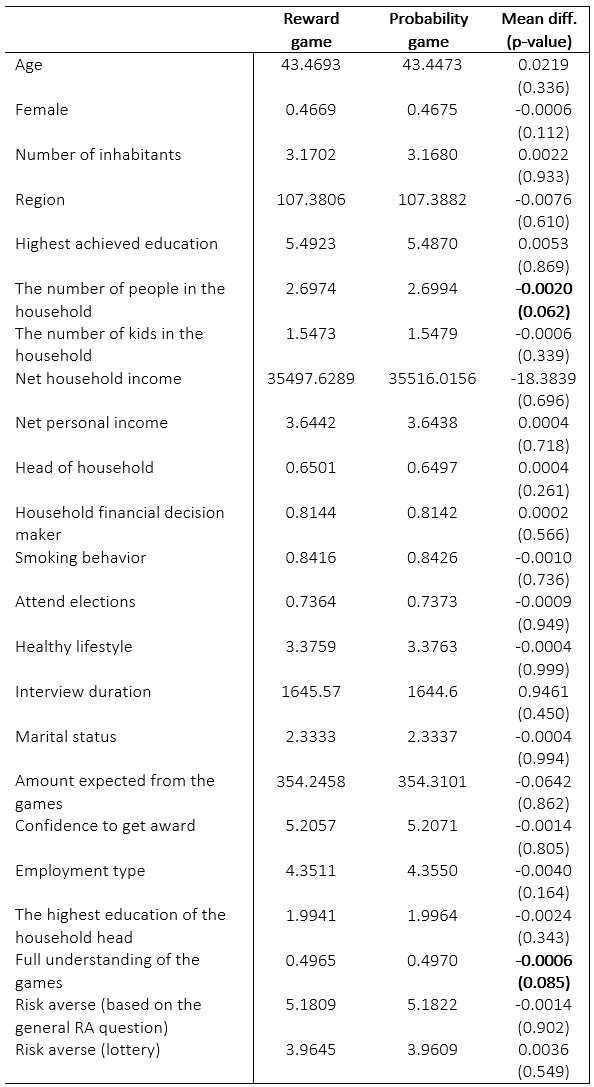}
    \caption{Sample descriptive statistics and randomization balance}
    \label{AppTab3:DescrStat}
    \end{center}
\end{table}

\begin{table}[H] 
\begin{center}
    \includegraphics[scale=0.6]{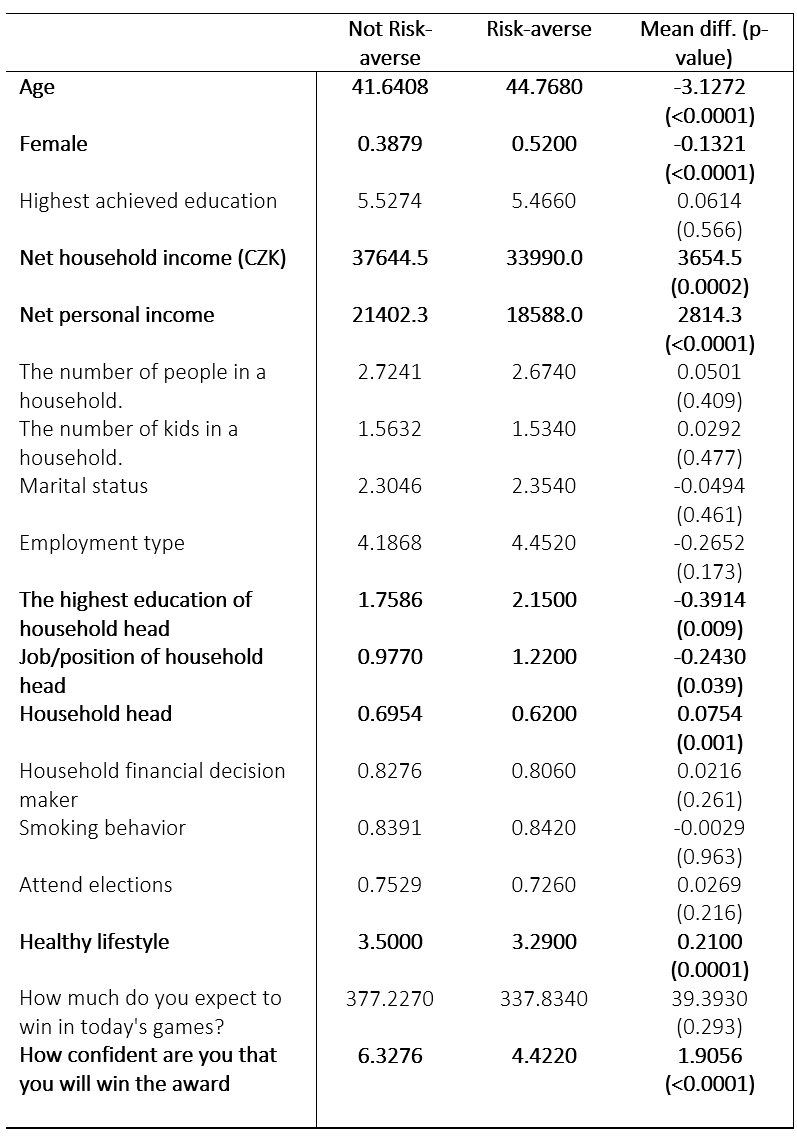}
    \caption{Comparison of people's characteristics by risk aversion}
    \label{AppTab4:RApeople}
\end{center}
\end{table}

In Table \ref{AppTab4:RApeople} we compare the difference between risk averse and not risk averse individuals. In line with existing literature, we observe that females and older people are significantly more risk averse. Risk averse participants have significantly lower personal income and live in a household with significantly lower household income. Risk averse participants do not differ in terms of highest achieved education, their employment type, marital status or household composition. While both groups expect on average the same return from the games, risk averse participants are significantly less confident about winning the awards.

\subsection{Robustness of the decision patterns}\label{Robust}

\begin{figure}[H] 
\begin{center}
    \includegraphics[scale=0.85]{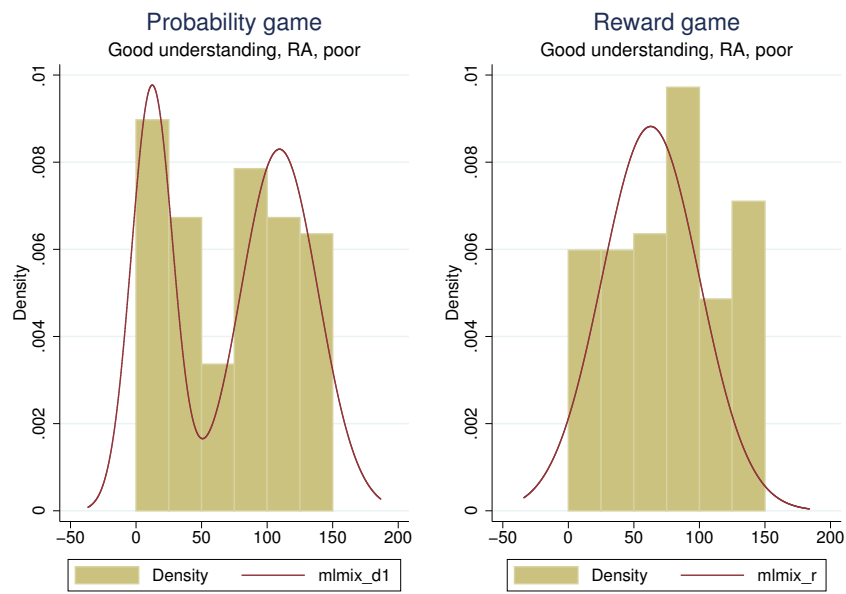}
    \caption{Investment decision of poor risk-averse people with good understanding}
    \label{AppFig5:FMM_poor_good}
\end{center}
\end{figure}

\begin{figure}[H] 
\begin{center}
    \includegraphics[scale=0.85]{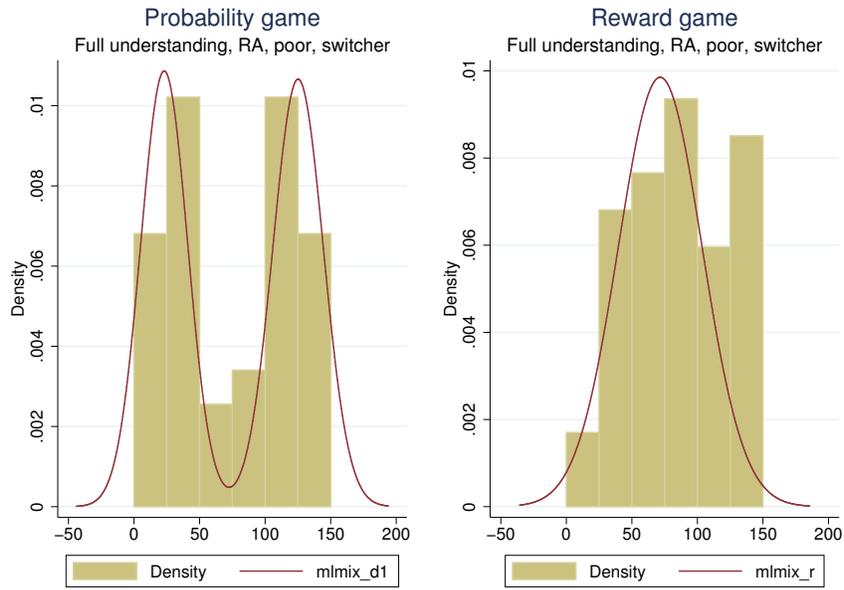}
    \caption{Investment decision of poor risk-averse people with full understanding; excluding non-switchers}
    \label{AppFig6:FMM_nonsw_full}
\end{center}
\end{figure}

\begin{figure}[H] 
\begin{center}
    \includegraphics[scale=0.85]{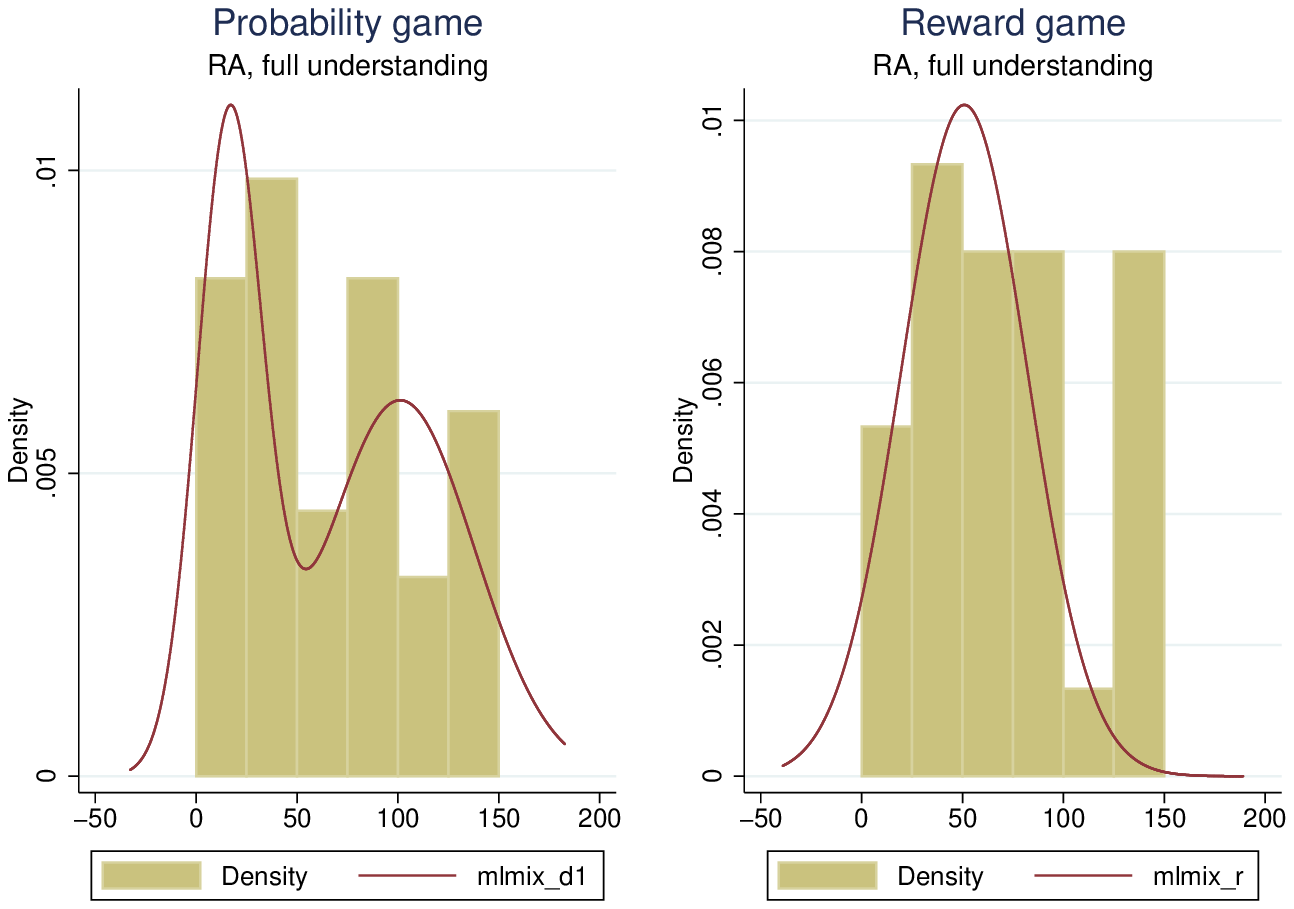}
\end{center}
    \caption{Investment decision of poor risk-averse people with full understanding, risk aversion measure based on Dohmen et al. (2010)}
    \label{AppFig7:FMM_RAdohmen}
\end{figure}

\begin{figure}[H] 
\begin{center}
    \includegraphics[scale=1]{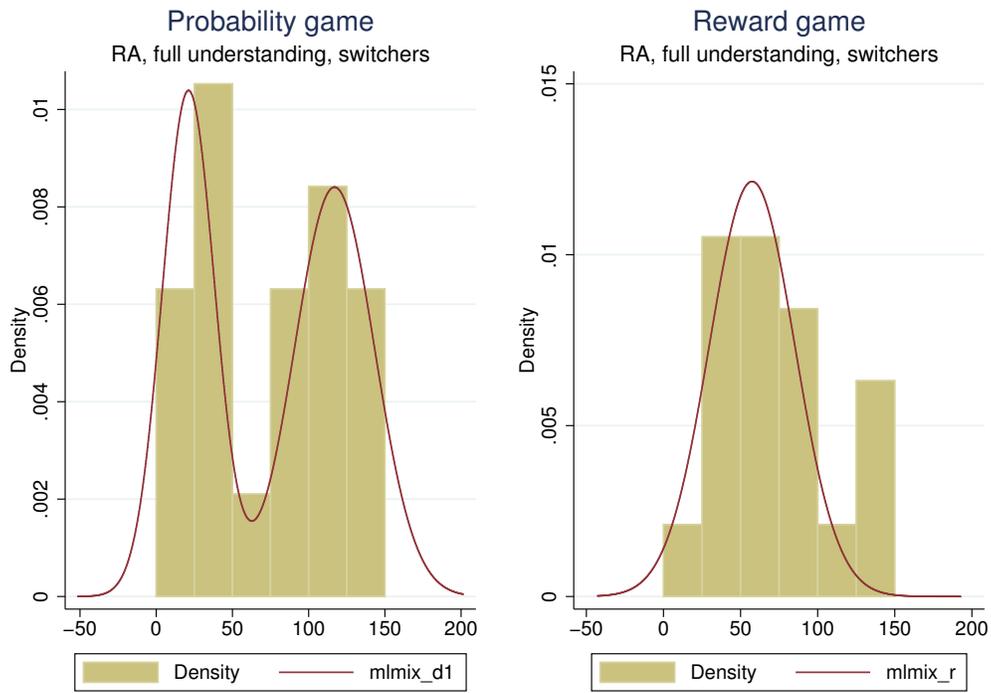}
\end{center}
    \caption{Investment decision of poor risk-averse people with full understanding excluding non-switchers, risk aversion measure based on Dohmen et al. (2010)}
    \label{AppFig8:FMM_RAdohmen_sw}
\end{figure}

In this subsection, we present supportive evidence to our main results. In Figures \ref{AppFig5:FMM_poor_good} up to \ref{AppFig8:FMM_RAdohmen_sw}, we look at the investment decisions of the poor risk-averse respondents with full or good understanding, using general measure of risk aversion (in Figures \ref{AppFig5:FMM_poor_good} and \ref{AppFig6:FMM_nonsw_full}) and the measure based on \cite{Dohmen2010} (in Figures \ref{AppFig7:FMM_RAdohmen} and \ref{AppFig8:FMM_RAdohmen_sw}). In all figures we see similar distribution shift from a bimodal distribution of investment decisions in the probability game to a unimodal distribution in the reward game. Once we exclude respondents who opted to invest equal amount in all three games (i.e., probability, reward as well as step-by-step game), the shift is even more pronounced (Figures \ref{AppFig6:FMM_nonsw_full} and \ref{AppFig8:FMM_RAdohmen_sw}).  

In the following tables, we present the results from probit regressions when people did not understand the setup of the games fully (\ref{AppTab6:probit_understanding}), if we control for various covariates (\ref{AppTab8:probit_covar}), and if we use different measures of risk aversion (\ref{AppTab7:probit_diffRAmeasures}). In all tables we see that risk averse individuals tend to invest significantly more in 0 CZK, 120 CZk, and 150 CZ, and significantly less in 60 CZK or 90 CZK.

\begin{table}[H]
	\centering
	\begin{threeparttable}		
		\begin{tabular}{lccccccccc}
			             & \\
             &       \multicolumn{2}{c}{Good understanding}\\ \hline\hline
             &    Risk averse&     Not risk averse\\ \hline
  Invested 0 CZK&    0.0179&   -0.0083\\ & {(0.013)} & {(0.018)} \\
 Invested 30 CZK&    -0.0149&   -0.0207\\ & {(0.018)} & {(0.022)} \\
 Invested 60 CZK&  \textbf{-0.0714}&    -0.0249\\  & {(0.023)} & {(0.027)} \\
 Invested 90 CZK&  -0.0446&    -0.0083\\  & {(0.028)} & {(0.032)} \\
Invested 120 CZK&  \textbf{0.0595}&      -0.0124\\  & {(0.022)} & {(0.028)} \\
Invested 150 CZK&  \textbf{0.0536}&  \textbf{0.0747}\\ & {(0.024)} & {(0.030)} \\
               n&     523&            328\\

		\end{tabular}
		 
	\end{threeparttable}
	\caption{Differences in investment probabilities between the two games for people with good (rather than full) understanding according to their risk-aversion}
	\label{AppTab6:probit_understanding}
\end{table}

\begin{table}[H] 

\begin{center}
    \centering 
    \begin{threeparttable}
        \begin{tabular}{lccccccccccccccc}
                         & \\
             &       \multicolumn{2}{c}{Model 2}& \multicolumn{2}{c}{Model 3} & \multicolumn{2}{c}{Model 4}\\ \hline\hline
             &    RA&     Not RA&       RA&      Not RA &       RA&      Not RA\\ \hline
Invested 0 CZK&    \textbf{0.0292}&  -0.0108&     \textbf{0.0253}&   -0.0107 & \textbf{0.0185} & -0.0037\\ & {(0.014)} & {(0.022)} & {(0.013)} & {(0.022)}& {(0.009)} & {(0.014)} \\
Invested 30 CZK&    -0.0118 &  -0.0431&     -0.0128&   -0.0398 & -0.0123 & \textbf{-0.0389}\\ & {(0.022)} & {(0.026)} & {(0.022)} & {(0.025)}& {(0.021)} & {(0.021)} \\
Invested 60 CZK&  \textbf{-0.0744}&   -0.0164 &  \textbf{-0.0744}& -0.0164&  \textbf{-0.0688}& -0.0073\\ & {(0.027)} & {(0.031)} & {(0.027)} & {(0.032)} & {(0.025)} & {(0.027)} \\
Invested 90 CZK&  \textbf{-0.0714}&   -0.0224 &  \textbf{-0.0721}&  -0.0225 & \textbf{-0.0755}&  -0.0240\\ & {(0.032)} & {(0.037)} & {(0.032)} & {(0.037)} & {(0.034)} & {(0.039)} \\
Invested 120 CZK&  \textbf{0.0465}&   -0.0162 & \textbf{0.0464}&  -0.0155 & \textbf{0.0462}&  -0.0154\\ & {(0.025)} & {(0.033)} & {(0.025)} & {(0.033)} & {(0.024)} & {(0.033)} \\
Invested 150 CZK&  \textbf{0.0848}& \textbf{0.1091}& \textbf{0.0839}&  \textbf{0.1106}& \textbf{0.0861}&  \textbf{0.1261}\\ & {(0.030)} & {(0.035)} & {(0.030)} & {(0.035)} & {(0.030)} & {(0.037)} \\
               n&                474&               366&          474&            366 &                474&               366\\

        \end{tabular}
    \end{threeparttable}
\end{center}
     \begin{minipage}{\linewidth}\footnotesize 
         Note: RA stands for risk averse. The estimated coefficients represent the marginal values and measure how much more likely are people invest in the amount if they play probability game compared to the return game. The numbers in brackets are standard errors. In Model 2 we also control for the order of the games, in model 3 we control for the order and household income, and in model 4 controls for all covariates.
    \end{minipage}
    \caption{Differences in investment probabilities between the two games, by different covariates}
    \label{AppTab8:probit_covar}

\end{table}

\begin{table}[H] 

\begin{center}
    \centering 
    \begin{threeparttable}
        \begin{tabular}{lccccccccccccccc}
                         & \\
             &       \multicolumn{2}{c}{Model 5}& \multicolumn{2}{c}{Model 6} & \multicolumn{2}{c}{Model 7}\\ \hline\hline
             &    Simple &     Covar &    Simple &     Covar   &    Simple &     Covar\\ \hline
Invested 0 CZK&    \textbf{0.0295}&  \textbf{0.0185}&    {0.0167}&   0.0121 & \textbf{0.0343} & 0.0028\\ & {(0.014)} & {(0.009)} & {(0.014)} & {(0.011)}& {(0.020)} & {(0.014)} \\
Invested 30 CZK&    -0.0127 &  -0.0123&     -0.0250&   -0.0279 & -0.0147 & {-0.0170}\\ & {(0.022)} & {(0.021)} & {(0.019)} & {(0.017)}& {(0.026)} & {(0.022)} \\
Invested 60 CZK&  \textbf{-0.0759}& \textbf{-0.0688} &  \textbf{-0.0667}& \textbf{-0.0169}&  \textbf{-0.0686}& \textbf{-0.0598}\\ & {(0.028)} & {(0.025)} & {(0.023)} & {(0.022)} & {(0.028)} & {(0.025)} \\
Invested 90 CZK&  \textbf{-0.0717}&   \textbf{-0.0755} &  \textbf{-0.0472}&  \textbf{-0.0495} & {-0.0196}&  -0.0198\\ & {(0.032)} & {(0.034)} & {(0.026)} & {(0.027)} & {(0.034)} & {(0.037)} \\
Invested 120 CZK&  \textbf{0.0464}&  \textbf{0.0462} & {0.0083}&  0.0079 & {-0.0049}&  -0.0007\\ & {(0.025)} & {(0.024)} & {(0.022)} & {(0.021)} & {(0.029)} & {(0.024)} \\
Invested 150 CZK&  \textbf{0.0844}& \textbf{0.0861}& \textbf{0.1139}&  \textbf{0.1159}& \textbf{0.0735}&  \textbf{0.0750}\\ & {(0.030)} & {(0.030)} & {(0.024)} & {(0.024)} & {(0.030)} & {(0.030)} \\
               n&    474&   366&    474&    366 &  474&   366\\

        \end{tabular}
    \end{threeparttable}
\end{center}
     \begin{minipage}{\linewidth}\footnotesize 
        
        Note: The estimated coefficients represent the marginal values and measure how much more likely are people invest in the amount if they play probability game compared to the return game. The numbers in brackets are standard errors. In Model 5 we use risk aversion measure based on the general risk aversion question, in Model 6 we use risk aversion measure inspired by \cite{Dohmen2010}, and in Model 7 we look at the very risk averse based on the \cite{Dohmen2010} measure. Regressions in columns titled Simple control for the treatment dummy only, regressions in columns titled Covar control also for other covariates.
    \end{minipage}
    \caption{Differences in investment probabilities between the two games, by various measures of risk aversion}
    \label{AppTab7:probit_diffRAmeasures}

\end{table}

\bigskip

\begin{figure}[H] 
%\begin{center}
    \includegraphics[scale=0.3]{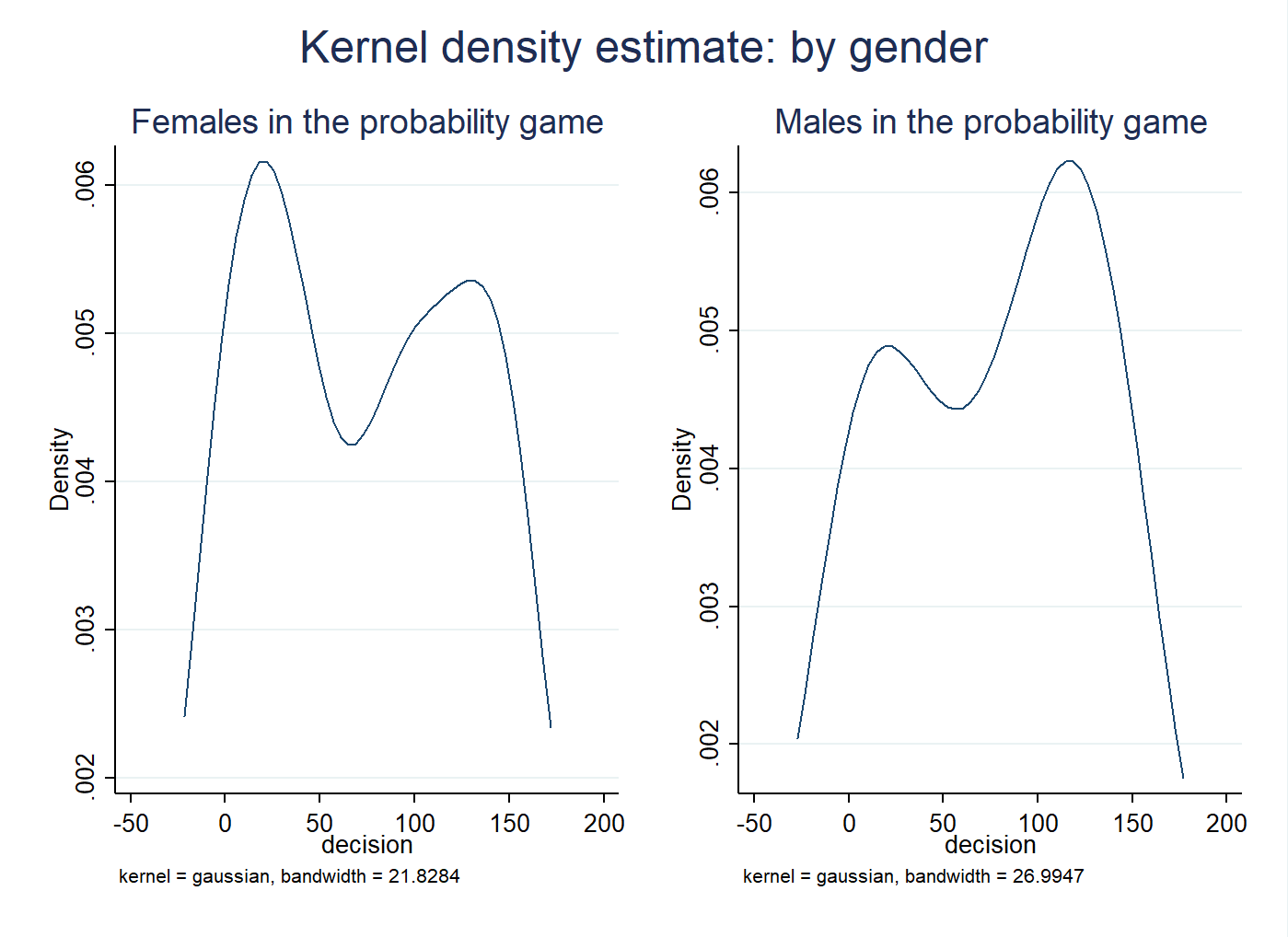}
 
\begin{minipage}{\linewidth}\small
          \footnotesize 
      Note: Smoothed density function of the distribution of investment choices of females and males in the probability game, estimated using a Gaussian kernel.
\end{minipage}
\caption{Comparison of density functions of the distribution of investment choices, by gender}
    \label{AppFig9:kdens_gender}
\end{figure}

In Figure \ref{AppFig9:kdens_gender} we present kernel density estimates of the investment choices in the probability game by gender. Females are known to be more risk averse. The figure suggests that in the probability game, females are more inclined to invest into minimum while males are more inclined to invest into maximum. 

\end{document}